\documentclass{aa}

\usepackage[colorlinks, citecolor=blue]{hyperref}
\usepackage{pdflscape}

\newcommand{\kms}{km\,${\rm s}^{-1}$}
\newcommand{\Msun}{M$_{\odot}$}
\newcommand{\vsini}{{$\varv\,\sin i$\,}}
\newcommand{\halpha}{H$\alpha$}

\defcitealias{Bodensteiner2020a}{Paper~I}
\usepackage{graphicx}

\usepackage{txfonts}

\begin{document}

   \title{The young massive SMC cluster NGC~330 seen by MUSE
          \thanks{Based on observations collected at the ESO Paranal
          observatory under ESO program 60.A-9183(A) and 0102.D-0559(A).}}
    \subtitle{II. Multiplicity properties of the massive-star population}

   \author{J. Bodensteiner\inst{1},
   H.~Sana\inst{1},
   C.~Wang\inst{2},
   N.~Langer\inst{2},
   L.~Mahy\inst{1,3},
   G.~Banyard\inst{1},
   A.~de Koter\inst{1,4},
   S.~E.~de~Mink\inst{5,4,6},
   C.~J.~Evans\inst{7},
   Y.~Götberg\inst{8},
   L.~R.~Patrick\inst{9},
   F.~R.~N.~Schneider\inst{10},
   F.~Tramper\inst{11}
   }

   \institute{Institute of Astronomy, KU Leuven, Celestijnenlaan 200D, 3001 Leuven, Belgium\\
              \email{julia.bodensteiner@kuleuven.be}
     \and
     Argelander-Institut für Astronomie, Universität Bonn, Auf dem Hügel 71, 53121 Bonn, Germany
     \and
     Royal Observatory of Belgium, Avenue Circulaire 3, B-1180 Brussel
     \and
     Astronomical Institute Anton Pannekoek, Amsterdam University, Science Park 904, 1098 XH, Amsterdam, The Netherlands
     \and
     Max-Planck-Institut für Astrophysik, Karl-Schwarzschild-Straße 1, 85741 Garching, Germany
     \and
     Harvard-Smithsonian Center for Astrophysics, 60 Garden St., Cambridge, MA 02138, USA
     \and
     UK Astronomy Technology Centre, Royal Observatory, Blackford Hill, Edinburgh, EH9 3HJ, UK
     \and
     The observatories of the Carnegie institution for science, 813 Santa Barbara Street, Pasadena, CA 91101, USA
     \and
     Departamento de F\'{\i}sica Aplicada, Facultad de Ciencias, Universidad de Alicante, Carretera de San Vicente s/n, E03690, San Vicente del Raspeig, Spain
     \and
     Heidelberger Institut für Theoretische Studien, Schloss-Wolfsbrunnenweg 35, 69118 Heidelberg, Germany
     \and
     Institute for Astronomy, Astrophysics, Space Applications $\&$ Remote Sensing, National Observatory of Athens, P. Penteli, 15236 Athens, Greece
    }
   \date{Received xx Month Year; accepted xx Month Year}

  \abstract
   {Observations of massive stars in open clusters younger than $\sim$8\,Myr have shown that a majority of them are in binary systems, most of which will interact during their life. While these can be used as a proxy of the initial multiplicity properties, studying populations of massive stars older than $\sim$20\,Myr allows us to probe the outcome of such interactions after a significant number of systems have experienced mass and angular momentum transfer and possibly even merged.}
   {Using multi-epoch integral-field spectroscopy, we aim to investigate the multiplicity properties of the massive-star population in the dense core of the $\sim\,40$\,Myr-old cluster NGC~330 in the Small Magellanic Cloud in order to search for possible imprints of stellar evolution on the multiplicity properties.}
   {We obtained six epochs of VLT/MUSE observations operated in wide-field mode with the extended wavelength setup and supported by adaptive optics. We extract spectra and measure radial velocities for 
   stars brighter than $m_\mathrm{F814W}=19$. 
   We identify single-lined spectroscopic binaries through significant RV variability with a peak-to-peak amplitude larger than 20\,\kms. We also identify double-lined spectroscopic binaries, and quantify the observational biases for binary detection. In particular, we take into account that binary systems with similar line strength are difficult to detect in our data set.}
   {The observed spectroscopic binary fraction among the stars brighter than $m_\mathrm{F814W}=19$ (approx.\ 5.5\,\Msun{} on the main sequence) is $f_\mathrm{SB}^\mathrm{obs} = 13.2 \pm 2.0\%$. Considering period and mass ratio ranges from $\log(P)$=0.15 - 3.5 (about 1.4 to 3160 d), and $q=0.1 - 1.0$, and a representative set of orbital parameter distributions, we find a bias-corrected close binary fraction of $f_\mathrm{cl}= 34^{+8}_{-7}$\%. This fraction seems to decline for the fainter stars, which indicates either that the close binary fraction drops in the B-type domain, or that the period distribution becomes more heavily weighted towards longer orbital periods. We further find that both fractions vary strongly in different regions of the color-magnitude diagram that corresponds to different evolutionary stages. This probably reveals the imprint of the binary history of different groups of stars. In particular, we find that the observed spectroscopic binary fraction of Be stars ($f_\mathrm{SB}^\mathrm{obs} = 2\pm2\%$)
   is significantly lower than the one of B-type stars ($f_\mathrm{SB}^\mathrm{obs} = 9\pm2\%$).}
   {In this work we provide the first homogeneous RV study of a large sample of B-type stars at a low metallicity ([Fe/H]$\lesssim$-1.0). The overall bias-corrected close binary fraction (log(P) < 3.5 d) of the B-star population in NGC~330 is lower than the one reported for younger Galactic and LMC clusters in previous works. More data are, however, needed to establish whether the observed differences result from an age or a metallicty effect.}

    \titlerunning{The young massive SMC cluster NGC 330 seen by MUSE. II.}
   \authorrunning{Bodensteiner et al.}

   \maketitle
%

\section{Introduction}\label{Sec:intro}
Observations have shown that a majority of OB stars, which are stars with initial masses larger than 8\,\Msun, are members of binary or higher-order multiple systems \citep[see e.g.,][]{Abt1990, Shatsky2002, Sana2012, Rizzuto2013, Sana2014, Kobulnicky2012, Dunstall2015, Moe2017}. A high fraction of those will interact during the course of their life \citep{Sana2012, Sana2013, Dunstall2015} which will drastically alter the subsequent evolution of both binary components \citep{Paczynski1967, Podsiadlowski1992}.

In contrast to the evolutionary pathway of single stars, interactions in binary systems can lead to a multitude of post-interaction pathways. This depends most strongly on the initial mass of the primary, and the initial orbital properties of the system, especially on the initial mass ratio and orbital period \citep{Podsiadlowski1992, Langer2012}. Very tight binaries are thought to undergo contact evolution and eventually merge 
\citep[see e.g.,][]{Pols1994, Wellstein2001, deMink2009a}. In the case of stable mass transfer via Roche lobe overflow (RLOF) 
the primary is stripped of its envelope \citep{Kippenhahn1967}. The companion accretes material and angular momentum, and is spun up to close to critical spin rates, depending on the amount of material transferred, and the ability of tides to spin it down \citep[e.g.,][]{Pols1991, deMink2013}.

Binary interactions thus lead to a variety of binary interaction products (BiPs) that can have very different observational characteristics. If originally part of a coeval star-formation event, for example in a single-starburst stellar cluster, mass gainers and merger products can appear as rejuvenated, single stars, some of which are situated above the cluster turnoff (TO). Observationally, stars above the TO are called blue stragglers \citep[e.g.,][]{Ferraro1997} or red stragglers if the BiP evolves to the red supergiant phase \citep{Britavskiy2019, Beasor2019}. In the case of RLOF, the secondary is spun-up and can be observed as a rapidly rotating star \citep{Pols1991} whereas the stripped core of the original primary is hard to detect  \citep{Wellstein2001, Gotberg2017}. If the remaining mass of the stripped star is $\lesssim$ 1\Msun, it will evolve into a subdwarf O or B star \citep[sdO/sdB; see e.g.,][]{Gies1998, Heber2009, Peters2013, Wang2017, Wang2018, WangL2021, Chojnowski2018, Gotberg2018}. If the core of the stripped star is massive enough, it will end in a core-collapse supernova, leaving behind a neutron star or black hole. Most systems will unbind as result of the natal kick of the compact object creating a walkaway or runaway star \citep[e.g.,][]{Eldridge2011, Renzo2019}. Systems that remain bound are possible progenitors for X-ray binaries \citep{Reig1997, Haberl2016}, depending on their configuration and evolutionary stage.

Given that a high fraction of massive stars interact with a companion during their life time the number of BiPs in a given population of massive stars is expected to become significant \citep{deMink2014, Britavskiy2019, Wang2020}. Theoretical computations have  shown that the fraction of BiPs as well as the currently observed binary fraction varies as a function of the age of the population \citep{Schneider2015}.

In order to gain insight in initial multiplicity properties, the binary properties of massive stars in young clusters both in our Galaxy as well as in the Large Magellanic Cloud (LMC) have been well-studied through spectroscopy \citep[e.g.,][]{Kobulnicky2012, Sana2012, Sana2013}.
Such spectroscopic studies are typically sensitive to systems with periods up to about 10 years, with a decreasing sensitivity above a couple of years. In this paper, we adopt the following nomenclature: systems detected through spectroscopy are called spectroscopic binaries $f_\mathrm{SB}^\mathrm{obs}$ while the corresponding intrinsic binary population, after bias correction, are called close binaries $f_\mathrm{cl}$. Unless specified otherwise, we adopt a maximum period of 10$^{3.5}$\,days to allow direct comparison with previous studies.

Investigating O stars in young open galactic clusters, \citet{Sana2012} found a bias-corrected close binary fraction of $f_{cl} = 69 \pm 9\%$. This is similar to the bias-corrected close binary fraction of OB stars in Cygnus OB2 of $\sim$\,55\% \citep{Kobulnicky2014}. The bias-corrected close binary fraction of young O-type stars in the 30 Dor region in the LMC was found to be $51 \pm 4\%$ \citep{Sana2013, Almeida2017}. Focusing on young B-type stars, \citet{Raboud1996} estimated a lower limit of 52\% on the observed close binary fraction in the young Galactic cluster NGC~6231.
In a recent study, \citet{Banyard2021} revisit the B-star population of NGC~6231 and find a bias-corrected close binary fraction of $f_{cl} = 44 \pm 6$\%. In the LMC, \citet{Dunstall2015} investigated B stars in the young cluster 30 Dor and found $f_{cl}$\,=\,58\,$\pm$\,11\%.

Homogeneous studies of the multiplicity of massive stars in intermediate-age clusters are, however, still lacking. In clusters of $\sim$20-40 Myrs, O stars will have evolved already off the main sequence (MS) and B-type stars now dominate the cluster TO \citep{Ekstrom2008,Brott2011}. Furthermore, a significant fraction of the initial binary systems will have had time to interact so that the number of BiPs in these clusters is expected to be significant \citep{Schneider2015}.

Investigating the current multiplicity properties of intermediate-age clusters as well as the frequency and characteristics of BiPs thus provides an ideal laboratory to test the current theories of binary evolution, in particular the physics controlling the interactions. One such intermediate-age cluster is NGC~330 in the Small Magellanic Cloud (SMC), reported to be between 25 and 45 Myrs old \citep[e.g.,][]{Sirianni2002, Milone2018, Patrick2020}.
While massive stars in the SMC field are reported to have a metallicity of [Fe/H]$=-0.7$ \citep[e.g.,][]{Luck1998, Korn2000, Keller2006}, NGC~330 was suggested to have an even lower metallicity of [Fe/H]$\lesssim-1.0$ \citep{Grebel1992, Gonzalez1999, Piatti2019}.
Due to the high stellar density in NGC~330, the cluster core could not be studied spectroscopically with past instrumentation. The advent of adaptive optics (AO) at the Multi-Unit spectroscopic Explorer (MUSE) in 2017 changed the situation \citep{Bacon2010}.

In \citet[][hereafter \citetalias{Bodensteiner2020a}]{Bodensteiner2020a}, we focused on a single epoch of AO-supported MUSE observations of the SMC cluster NGC~330 (epoch \#4, see Table\,\ref{Tab:obs}).  \citetalias{Bodensteiner2020a} described the observations, data reduction and sky subtraction. Furthermore, we gave a detailed explanation on how spectra for individual stars were extracted from the MUSE data cube using PSF-fitting. Using an automated spectral classification based on equivalent widths of He lines, we assigned spectral types to more than 200 stars. We derived a total cluster mass of $88^{+17}_{-18} \times 10^3\,$\Msun\, and an age of the cluster core of 35-40 Myrs.

NGC~330 was previously reported to contain a large number of Be stars \citep[see e.g.,][]{Feast1972, Grebel1996, Keller1999, Martayan2007b}. Based on HST photometry, \citet{Milone2018} find that the color-magnitude diagram (CMD) of NGC~330 shows a split MS and a large population of \halpha\, emitting stars, interpreted as Be stars. Their colors indicate that they are cooler than the bulk of MS stars, which is why these stars predominantly populate a region offset off the MS region (see their figure 10).
Using the MUSE observations, we confirmed that stars in this cooler region in the CMD are indeed Be stars. We furthermore found that the overall observed Be star fraction in the cluster core is 32$\pm$3\%. Around and above the cluster turnoff, it rises to $\sim 50 \pm 10 \%$, as already indicated by \citet{Milone2018}.

Be stars are observationally identified as B-type stars with strong Balmer-line emission \citep{Jaschek1981}. As this observational definition includes several objects of different physical nature (i.e., Herbig Ae/Be stars, B-type supergiants, interacting binaries), the term "classical Be stars" is used to describe rapidly-rotating non-supergiant B-type stars with a circumstellar decretion disk giving rise to the characteristic line emission \citep{Rivinius2013}. The origin of the rapid rotation of classical Be stars, which is thought to be close to the critical breakup velocity, is still debated \citep{McSwain2005}. According to the proposed single-star channels, classical Be stars could be born as rapid rotators \citep{Bodenheimer1995} or spin-up towards the end of their MS lifetime \citep[see e.g.,][]{Ekstrom2008, Hastings2020}. Another proposed formation channel interprets classical Be stars as mass gainers in previous binary interactions \citep{Pols1991, deMink2013, Shao2014, Bodensteiner2020b}. A way to test the relative contributions of these channels is to investigate the multiplicity properties of classical Be stars: if the single-star channels dominate their formation, Be stars are expected to have similar binary properties than normal B-type stars. If the binary channel, however, is the dominating one, Be stars are not expected to be in close binary systems with MS companions but rather have stripped or evolved companions -- which are hard to detect observationally -- or be runaways \citep{Bodensteiner2020b}.

Here, we extend our analysis to study the multiplicity properties of NGC~330 using six epochs of MUSE observations. This paper is organised as follows.  Section\,\ref{Sec:ObsAndSpec} describes the multi-epoch observations, their reduction and the extraction of individual spectra.  Section\,\ref{Sec:RVmeas} focuses on the measurement of radial velocities (RVs) while we discuss the correction for observational biases in Sect.\,\ref{subsec:obsbias}. We derive the spectroscopic binary fraction of the overall massive-star population of NGC~330 in Sect.\,\ref{Sec:binarity} and discuss individual groups of stars of interest. Finally, in Sect.\,\ref{Sec:discussion} we discuss our results while the conclusions and an outlook to future work are given in Sect.\,\ref{Sec:conclusion}.

\section{Observations, sample, and extraction of spectra}\label{Sec:ObsAndSpec}
\subsection{Observations and data reduction}\label{Subsec:obs}
Six epochs of observations of NGC~330 were obtained with MUSE between August 2017 and December 2018. The six epochs, which were split into four dither positions of 540 seconds each, were spread out over approximately 1.5 years. A journal of the observations with seeing conditions is given in Table\,\ref{Tab:obs}. Additionally, the average full width at half maximum (FWHM) of the point spread function (PSF) as measured by fitting 2D Gaussian profiles on several bright and isolated targets in the reduced and collapsed white-light image is provided for each epoch. It gives an indication of the final image quality after data reduction, taking into account the effect of the AO.  \newline
\begin{table}
    \caption{\label{Tab:obs} Journal of the MUSE observations of NGC~330. The first column assigns a number to each epoch while columns two and three give the date and modified Julian date (MJD) at the start of each observation. The last two columns provide the  seeing during the observations as measured by the Paranal Differential Image Motion Monitor (DIMM), and the average FWHM of the PSF measured in the reduced data cubes.}
    \begin{tabular}{lcccc} \hline \hline
        Nr. & date & MJD & Seeing & FWHM \\
         &      & [d] &   ["]  &  ["] \\ \hline
        1 & 2017-08-15 & 57980.22523264 & 0.4 - 0.9 & 0.7 - 0.8 \\
        2 & 2017-09-18 & 58014.12963530 & 1.2 - 1.9 & 0.8 - 0.9 \\
        3 & 2018-11-06 & 58428.09317181 & 0.8 - 1.1 & 0.8 - 1.0 \\
        4 & 2018-11-19 & 58441.07863659 & 0.4 - 0.7 & 0.6 - 0.7 \\
        5 & 2018-11-22 & 58444.04752694 & 0.6 - 0.8 & 0.7 - 0.8 \\
        6 & 2018-12-18 & 58470.03819326 & 0.6 - 0.8 & 0.7 - 0.8 \\ \hline
    \end{tabular}
\end{table}
MUSE is made up of 24 individual spectrographs which, in wide-field mode (WFM), cover a field of view (FoV) of 1'$\times$1' with a spatial pixel sampling of 0.2'' in both spatial directions. At the estimated distance of $60\pm1$\,kpc, \citep[see e.g.,][]{Harries2003}, 1' corresponds to about 20\,pc and 0.2" to 0.07 pc or 1.4$\cdot$ 10$^4$\,AU. Observations in the extended wavelength mode cover the optical spectrum between 4600\,$\AA$ and 9300$\AA$. When using the AO mode, the region between 5780\,$\AA$ and 5990\,$\AA$ is blocked to avoid stray light from the laser guide stars. The spectral resolving power $\lambda/\Delta\lambda$ varies from 1700 at $4600\,\AA$ to 3700 at $9300\,\AA$.

All epochs were reduced as described in \citetalias{Bodensteiner2020a}. In short, we used the standard ESO MUSE pipeline v2.6\footnote{\url{https://www.eso.org/sci/software/pipelines/muse/}} to apply bias and dark subtraction, flat fielding, wavelength and illumination correction, telluric correction and  flux calibration. A sky subtraction was performed by taking the darkest pixels in each observation to create an average sky spectrum that was then subtracted from each pixel in each data cube.

\subsection{Extraction of spectra}
The extraction of spectra from the fully reduced MUSE data cubes is done via PSF-fitting. In order to take into account the crowding in the field, we fit the PSF of multiple, close-by sources simultaneously to limit contamination. For this purpose, we need a catalogue of accurate stellar positions and brightness, preferably from an instrument with higher spatial resolution, which is then used as input for the PSF fitting.

In contrast to the procedure adopted in \citetalias{Bodensteiner2020a}, we use the photometric catalogue provided by the GALFOR project\footnote{\url{http://progetti.dfa.unipd.it/GALFOR/MCs/NGC0330.html}}, described in \citet{Milone2018}, as target input catalogue. The catalogue, which is based on Hubble space telescope (HST) wide field camera 3 (WFC3) observations taken in 2015 and 2017, contains coordinates as well as magnitudes in different filters (i.e., F225W, F336W, F656N, and F814W). No magnitudes are available for three of the red supergiants (RSG) and for one blue supergiant (BSG). Additionally, a handful of bright sources fall into the gap between the individual WFC3 chips. We estimated their position and magnitude by comparison to similarly bright stars and added them to the input catalogue manually. 

After aligning the HST catalogue with the MUSE coordinate system, stars in the MUSE FoV with F336\,W magnitudes brighter than 17.5 (Vega magnitudes, corresponding to stars with masses $\gtrsim$\,5.5\,\Msun{} on the MS) were selected for spectral extraction. For each of these 400 stars, the spectrum was extracted using a \textsc{python} routine based on PSF fitting (see \citetalias{Bodensteiner2020a} for more details). In short, the PSF of each star is fitted at each wavelength slice (i.e., 3701 times) with all stars closer than 12 pixel and brighter than 18.5 in F336W being fitted simultaneously. This ensures that the high level of crowding in the dense cluster core is taken into account as close-by sources are effectively deblended. The individual flux measurements per wavelength bin were then concatenated to a spectrum for each star.

When stars are too close to the edge of the FoV, the PSF-fitting routine fails. Spectra for these stars were extracted manually using QfitsView\footnote{\url{https://www.mpe.mpg.de/~ott/QFitsView/}}. The extraction in QfitsView is similar to aperture photometry, i.e. the flux of a star in each wavelength slice is determined by summing up the flux within a certain aperture (which was fixed to 3x3 pixels here). Given that the stellar density towards the edge of the MUSE FoV is lower, the sources here are more isolated and contamination by close-by sources is less likely.

Due to a small spatial offset between the epochs, the different observations do not cover the exact same region in the sky. As a result, some stars close to the edge of the MUSE FoV are not observed in every epoch and therefore have less than six epochs available. Furthermore, the seeing conditions varied (see Table\,\ref{Tab:obs}), which is why the quality of the individual spectra vary from epoch to epoch.

In total, we extracted at least four spectra for 400 stars brighter than m$_{F336W}=17.5\,\mathrm{mag}$. For 55 stars, mainly faint stars in crowded regions of the FoV, the quality of the extracted spectra i.e., the continuum signal-to-noise ratio (S/N) was $\lesssim 80$ and thus deemed insufficient for a subsequent analysis. Furthermore, there are five foreground stars in the FoV which are classified as such given that, while having a similar brightness to the B-type stars, they show the spectrum of K-M type star. The spectra of the remaining 340 stars were automatically normalised (see \citetalias{Bodensteiner2020a}). In total this provides us with normalised spectra for 324 B-type stars, ten RSGs and six BSGs.

\subsection{Spectral types}
Investigating the Balmer lines of the non-supergiant B-type stars for the presence of emission lines in at least one epoch, we identified 115 Be stars (i.e., 35\%). The majority of them are probably classical Be stars, that is rapidly rotating B-type stars with Balmer-line emission arising in a circumstellar decretion disk \citep{Rivinius2013}. As described in Sect.\,\ref{Sec:intro}, a similar spectral signature is, however, observed for young stellar objects, that is Herbig Ae/Be stars \citep{Waters1998}, as well as for magnetic stars \citep{Petit2013} or interacting binaries \citep{Horne1986}. Based on our current data, it is not possible to distinguish between those natures. Classical Be stars are, however, expected to populate a distinct region in the CMD \citep[see e.g.,][]{Milone2018}. Additionally, given the crowding of MUSE FoV, and despite the advanced spectral extraction technique used here, blending with another source and thus contamination in the H$\alpha$ line can not be fully excluded. In order to assess this, we count the number of faint sources (i.e., with F814W magnitudes > 18) with bright emission-line stars close by, and find that there are only 10 such objects.

For all other B-type stars, spectral types were assigned based on an equivalent width (EW) calibration of diagnostic spectral lines of standard stars, which was developed and described in \citetalias{Bodensteiner2020a}. As in \citetalias{Bodensteiner2020a}, Be stars as well as BSGs were classified compared to standard star spectra by visual inspection. For this we use the spectra extracted from epoch 4 given that it has the best image quality and therefore the lowest contamination.

Among the 340 remaining stars in our sample, there are 208 B stars, 116 Be stars, six BSGs and ten RSGs. The S/N per epoch varies between $\sim$80 for the fainter and $\sim$200 for the brighter MS stars while the BSGs have a typical S/N of about 300 at 6000$\AA$.

\section{Radial velocities and the multiplicity criteria}\label{Sec:RVmeas}
This section focuses on the multiplicity properties of the B and Be star populations as well as the BSGs. The multiplicity properties of the RSG population were studied extensively in \citet{Patrick2020} based on a more extended observational data set including also RSGs outside the MUSE FoV.

\subsection{RV measurements}\label{Subsec:rvmeas}
In order to assess the multiplicity status of each star, we measure the Doppler shift of a set of spectral lines in all available epochs. RVs are measured following the procedure described in \citet{Sana2013}, where Gaussian profiles are fitted to a set of spectral lines simultaneously for all epochs.
The shapes of the individual line profiles are kept constant across the epochs, and the different lines fitted for each epoch are required to be reproduced by the same RV. The method is thus more robust than line-by-line fitting, especially for epochs with poorer S/N than the average.

For the B and Be stars, we use a set of six \ion{He}{i} lines as standard line list: \ion{He}{i} $\lambda \lambda$ 4713.15, 4921.93, 5015.68, 5047.74, 6678.15, 7065.19 $\AA$\footnote{Rest wavelengths are taken from the NIST database.}. We first visually inspect all spectra of each star and, if necessary, reduce the number of lines used to a subset of the standard line list. Lines are excluded depending on the S/N of the spectra, which particularly affects the weak \ion{He}{i} lines at 5015 and 5047$\,\AA$. In addition, Be stars suffer from line infilling by emission which can significantly reduce the set of lines that can be used for RV measurements. On average, four \ion{He}{i} lines are used, but there is a handful of stars that only had one line available.

While spectra were extracted for 324 B and Be stars, RV measurements were only possible for 282 of them. This was mainly due to the low S/N in some of the spectra, either because the stars in question are too faint, or because they are close to significantly brighter sources, which adds to the photon noise of the target itself. Among the 282 B and Be stars, we have six epochs for 273 stars, five epochs for eight stars, and four epochs for the remaining star.

For the six BSGs in the sample we choose a different set of lines to measure RVs. Given that they are all early to late A-stars, their spectra do not show strong \ion{He}{i} lines. We therefore use a set of \ion{Fe}{ii} and \ion{Si}{ii} lines (i.e., \ion{Fe}{ii} $\lambda \lambda$ 4923.93, 5018.44$\,\AA$ and \ion{Si}{ii} $\lambda \lambda$ 5055.98, 6347.10, 6371.36$\,\AA$). For stars \#\,327 and \#\,586, we excluded the \ion{Fe}{ii} at 4923$\AA$ because it is blended with a weak \ion{He}{i} line. For the  late-A supergiant \#\,503, we only use the \ion{Si}{ii}  lines around 6350$\AA$ given the large number of contaminating spectral lines in the spectrum.

Our final catalogue, including coordinates, magnitudes, number of epochs, spectral types, RV measurements, the set of lines used, as well as the binary flag we assigned to them (see Sect.\,\ref{Subsec:bin_crits} and \ref{Subsec:sb2s}) and individual notes is given in Table\,\ref{tab:all_params}. This thus provides a large-scale homogeneous and self-consistent catalogue of RVs for almost 300 stars, the full version of which will be made available electronically at the CDS\footnote{CDS link}.

\subsection{Multiplicity criteria}\label{Subsec:bin_crits}
In order to assess whether measured RV variability is statistically significant, which allows us to classify stars as binary candidates, we adopt two multiplicity criteria \citep[see e.g.,][]{Sana2013, Dunstall2015, Patrick2019, Banyard2021, Mahy2021}. A star is classified as binary if at least one pair of RVs measured in different epochs satisfy the following two criteria simultaneously. The first criterion is
\begin{equation}\label{eq:one}
    \frac{|v_i - v_j|}{\sqrt{\sigma_i^2 + \sigma_j^2}} > 4.0 \quad ,
\end{equation}
where $v_i$ and $v_j$ are the individual RV measurements and $\sigma_i$ and $\sigma_j$ the respective RV errors at epoch $i$ and $j$. The confidence threshold of 4.0 is chosen so that the number of false positives is less than one given the sample size and number of epochs. As the RV errors are typically smaller for brighter stars, these are more likely to pass this criterion than fainter stars with larger RV uncertainties. This effect is, however, accounted for in the bias correction (see Sect.\,\ref{subsec:obsbias} and Table \ref{Tab:mag_pdetect}).

The second criterion takes into account that there are other processes that can lead to RV variability in massive stars, most importantly pulsations \citep{Aerts2009, Simon-Diaz2017} and wind variability \citep[see e.g.,][]{Fullerton1996}.
We therefore impose that two individual RV measurements have to deviate from one another by more than a minimum RV threshold:
\begin{equation}\label{eq:two}
    |v_i - v_j| > \Delta \mathrm{RV_{min}} \quad .
\end{equation}

\begin{figure}[t!]\centering
    \includegraphics[width=0.95\hsize]{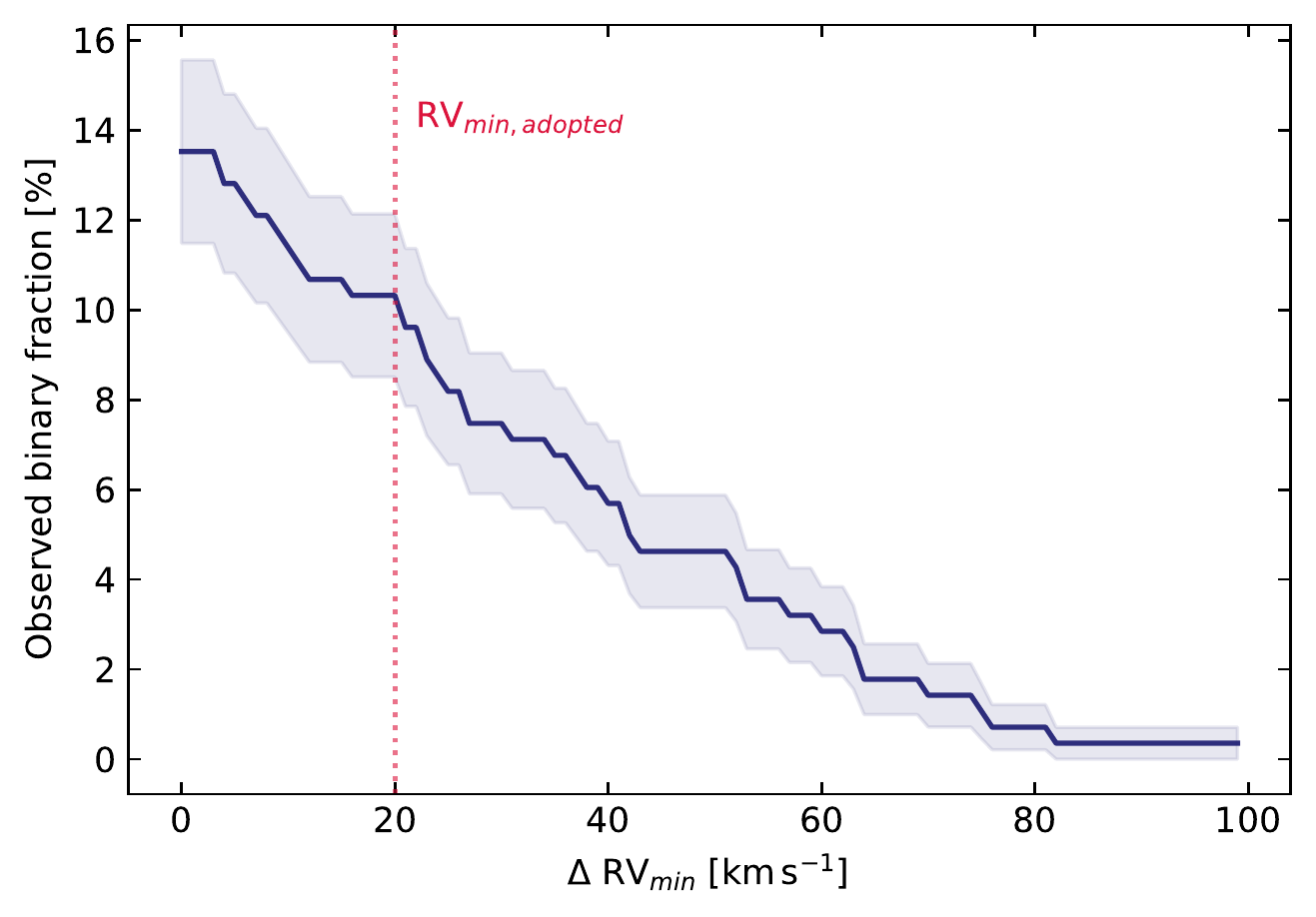}
    \caption{\label{Fig:rvmin} Observed close binary fraction for MS stars as a function of the adopted RV-variability threshold $\mathrm{\Delta RV_{min}}$. For each $\mathrm{\Delta RV_{min}}$ we compute the observed close binary fraction of the entire sample as well as the corresponding binomial error. The vertical dotted line gives the adopted threshold value $\mathrm{\Delta RV_{min}}=20$\,\kms\, indicated by the red line.}
\end{figure}

We investigate the impact of the choice of this threshold $\mathrm{\Delta RV_{min}}$  in Fig.\,\ref{Fig:rvmin}. It can be seen that the overall observed close binary fraction for MS stars is low, and that small changes in the threshold value will not change the results significantly. Here, we adopt $\mathrm{\Delta RV_{min}}=20$\,\kms, which is in line with similar studies \citep[see e.g.,][]{Sana2013, Dunstall2015, Banyard2021} and above the typical RV variations of B-type stars caused by pulsations \citep[e.g.,][]{Aerts2009}.

According to our two criteria, 24 out of the 189 B stars for which we could measure RVs show significant (Eq.\,\ref{eq:one}) and large (Eq.\,\ref{eq:two}) RV variability. Among the 93 Be stars with measured RVs, five pass these RV criteria, as well as one of the six BSGs. These stars are considered spectroscopic binaries in the following, even though the orbital period could not be constrained.

\subsection{Double-lined spectroscopic binaries}\label{Subsec:sb2s}
We identify additional binary candidates by visually inspecting the spectra of all the B and Be stars as well as BSGs, specifically looking for composite spectral signatures indicative of double-lined spectroscopic binaries (SB2s). These are not always flagged as binary by the applied RV criteria given that their spectral lines are typically not well represented by a single Gaussian fit resulting in inaccurate RV measurements with large error bars.

We focus on the above mentioned He\,I lines used for the RV measurements, as well as the Balmer lines. For Be stars, because their diagnostic spectral lines are often filled-in by emission, the visual classification of SB2s is particularly difficult. We find eight systems (\# 54, 55, 119, 156, 203, 212, 252, and 637) that show a composite spectrum indicative of two stars and are therefore classified as candidate SB2s. Two of those are Be stars while the remaining 6 are B-type stars. While they could be genuine SB2s, they could also be the line-of-sight superposition of two sources given the high stellar density in the cluster core. Given that the stars show line profile variations, which would not be expected for two single stars that are aligned by chance, this seems, however, unlikely. Our data quality does not allow us to derive accurate spectral types for the secondary components. Given that we detect them based on \ion{He}{i} lines, and that they contribute significantly to the flux in the optical, they are most likely B-type stars.

\section{Correction for observational biases} \label{subsec:obsbias}
In order to assess the intrinsic close binary fraction of the massive-star population in NGC~330, the observed spectroscopic binary fraction needs to be corrected for the observational biases. We specifically consider the following two detection biases. The first one is the chance to detect significant RV shifts above our detection threshold, depending on the orbital properties, temporal sampling and RV measurement accuracy. The second bias impacts SB2 binaries, or more precisely unidentified SB2 binaries. In these systems, the presence of the companion reduces the apparent Doppler shift of the primary star, henceforth reducing the chance of detecting large enough Doppler shift to meet our detection criteria.

\subsection{Single-line detection bias}\label{s:sb1-bias}

Following the approach described in \citet{Sana2013}, we estimate the binary detection sensitivity of our observing campaign by simulating a population of binary systems. Corresponding to the number of stars in our sample we measured RVs for, we simulate 280 binary systems taking into account our observational setup (i.e., the time coverage and number of observational epochs). For each system, we randomly select the initial mass of the primary $M_1$ from a Salpeter initial mass function \citep{Salpeter1955} between 4.5 and 8~\Msun. The choice of the range of primary masses is guided by the mass distribution of the MS stars in our sample, estimated in \citetalias{Bodensteiner2020a}. We then pair them with a companion by randomly drawing the mass ratio $q$, the orbital period $P$, and the eccentricity $e$ from parent distributions of orbital parameters. We further adopt a random orientation of the orbital plane in 3D space and a random reference time in the ephemeris. We first assume a flat period distribution in $\log\,P$, ranging from 0.15 to 3.5 (i.e., $P$ from 1.4 to 3160 days), a flat mass-ratio distribution between $q = 0.1$ and 1, and an eccentricity distribution between $e = 0$ and 0.9 which is proportional to $\sqrt{e}$, with a simple circularisation correction for short periods. These parameter ranges are adopted to allow for a direct comparison with other works. The choice of the underlying parent distributions of orbital parameters will be discussed in Sect.~\ref{s:pqe_distr}.

\begin{figure}\centering
    \includegraphics[width=0.98\hsize]{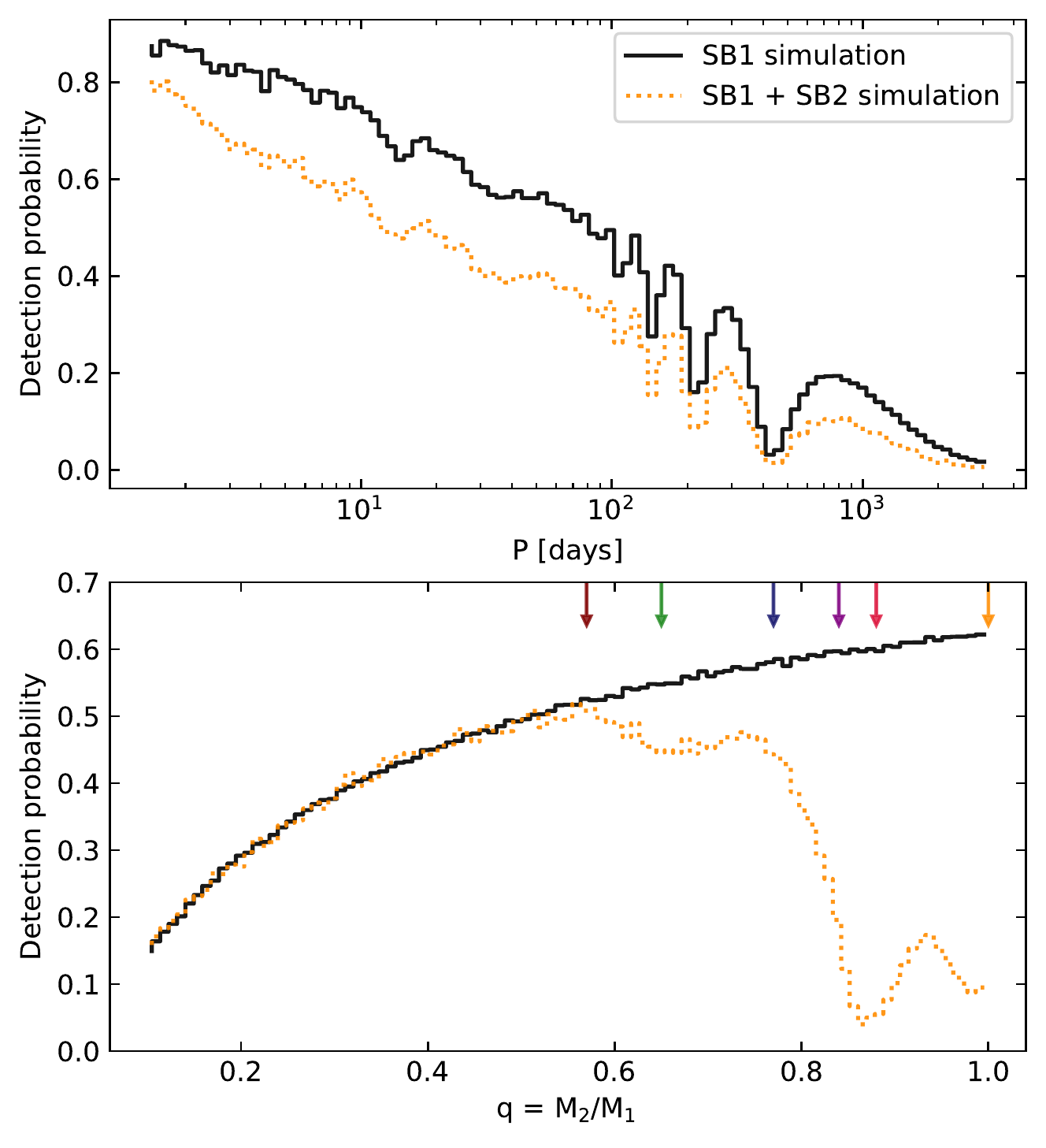}
    \caption{\label{Fig:bias_corr} Binary detection probability simulations following our observational setup. The top panel shows the detection probability as a function of the orbital period $P$ (computed assuming a flat mass-ratio distribution); the bottom panel as a function of mass ratio $q$ (computed assuming a flat $\log P$ distribution). In both panels, the black line shows our first estimate of the binary detection probability taking into account the SB1 bias, while the orange line shows the detection probability also including the reduced sensitivity to detect near equal-mass binaries. The colored arrows in the bottom panel correspond to the mass ratio for which we compute the SB2 bias (see Fig.\,\ref{Fig:rvmeas_results}).}
\end{figure}

After assigning RV errors, taken from the observations, we apply the two multiplicity criteria described in Eqs.\,\ref{eq:one} and \ref{eq:two}. By repeating the simulation of 280 binary systems 10\,000 times\footnote{The number of simulations is chosen so that the associated statistical uncertainty due to the number of simulations is approximately five times smaller than the one due to the sample size.}, we compute the probability of detecting these systems as binaries with our observing campaign. This thus gives an estimate of the sensitivity of our observation, and provides us with a correction factor that is multiplied to the observed binary fraction in order to estimate the intrinsic close binary fraction. A second run is then perform adopting the intrinsic binary fraction as input to allow for a proper estimate of uncertainties due to the binary population sample size. We find that the overall detection probability over the period range of 1.4 to 3160 days is $\sim45$~\%. As shown in Fig.\,\ref{Fig:bias_corr}, the detection probability depends strongly on the period and mass ratio: while the detected probability is almost 90\%\ for the shortest period systems, it drops below 50\%\ for periods larger than $\sim$100\,days. Two-dimensional bias correction plots are shown in the Appendix \ref{App:bias}.

\subsection{Double-line detection bias}\label{s:sb2-bias}
SB2 systems for which both components show similar line strengths suffer from an additional detection bias: the apparent Doppler shift of blended spectral lines can be reduced or even masked entirely. This is caused by the fact that the superposition of two similar line profiles is hard to distinguish from a single line profile if the RV separation is not large enough and the variation of the (photo-)center of this presumably single line is reduced compared to the true orbital Doppler shift \citep[e.g.,][]{Sana2011b}. This SB2 bias is lifted as soon as one can identify the SB2 nature of the system and can use a double-line profile fit to measure the RVs of both components simultaneously. As such, this bias is stronger for broader lines (either due to rotation or spectral resolution) and lower S/N.

Given the low resolving power of the MUSE spectra (i.e., between 1700 and 3700), we perform more detailed simulations in order to investigate the minimum RV separation at which their spectral lines are deblended and the two stars would be classified as SB2s. This allows us to quantify the bias that impacts the RVs measurements of unidentified SB2 systems. We use synthetic spectra from the \textsc{tlusty} B-star grid computed for SMC metallicity \citep{Lanz2007} and select atmospheric models corresponding to spectral types B2 to B5, following the relation between stellar parameters and spectral types from \citet{Silaj2014}. We perform this simulation for a  B2~V single star, and a set of binary systems with a B2~V primary, and a mass ratio $q$ ranging from 0.5 to 1.0. We furthermore broaden the spectra for two rotational velocities \vsini of 100 and 200 \kms, which is typical for stars in the considered mass range. We shift the stars spectra in steps of 10\,\kms, with respect to one another, co-add them accounting for their light ratio, degrade the simulated spectra to the MUSE resolution and a binning of 1.25\,$\AA$ (corresponding to the binning of our reduced spectra), adopt the MUSE wavelength coverage and a typical S/N ratio corresponding to the observations.

\begin{figure}\centering
    \includegraphics[width=0.99\hsize]{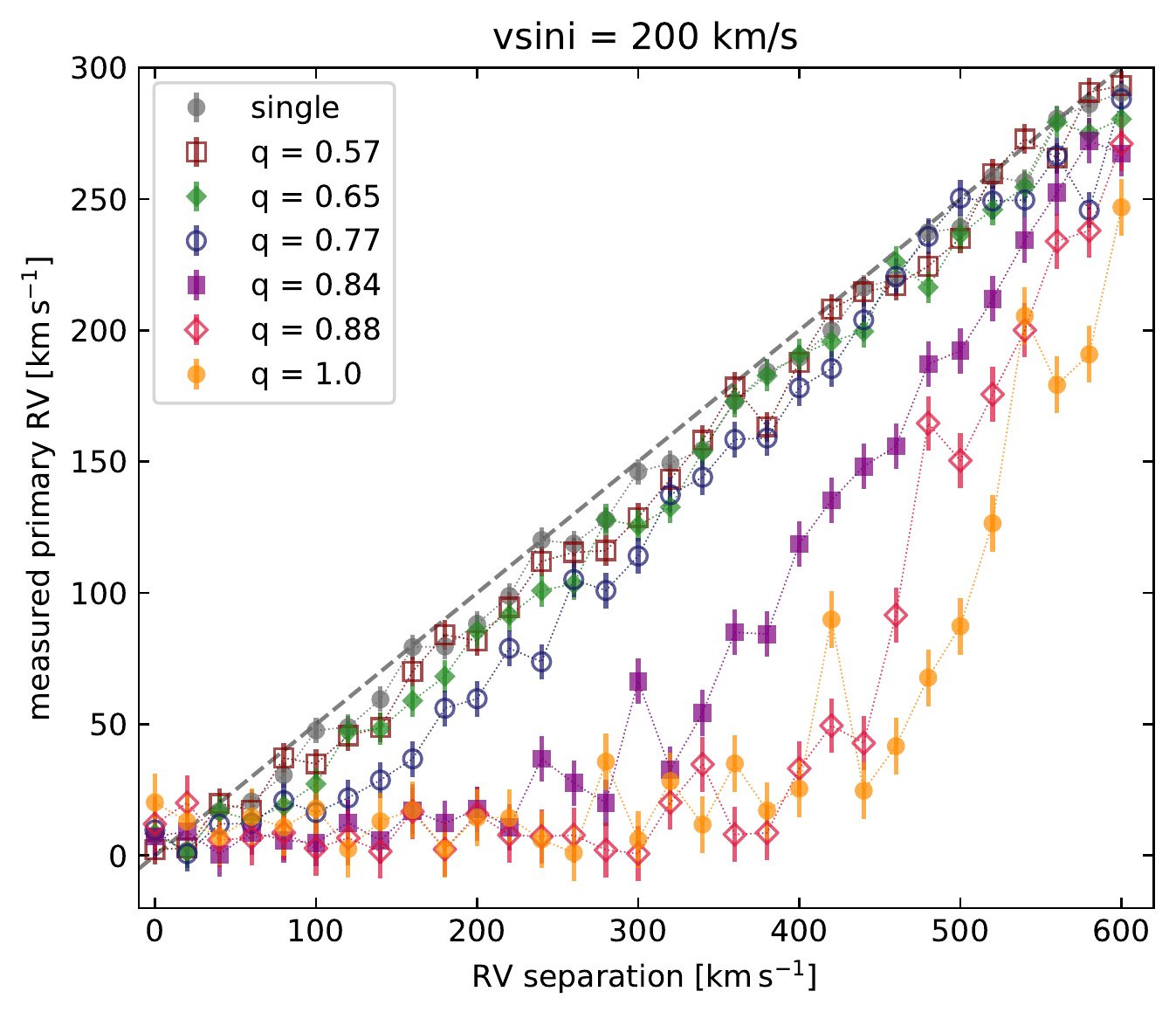}
    \caption{\label{Fig:rvmeas_results} Measured primary RVs for simulated spectra of one single B2~V star, and SB2 binary systems with a B2~V primary and six different mass ratios as indicated in the legend. A rotational velocity of \vsini = 200\,\kms\,is assumed for all stars. The horizontal axis gives the assumed RV difference between the two components. The dashed line indicates the true velocity of the primary star.}
\end{figure}

We then measure RVs with the same method as for our sample stars. Fig.\,\ref{Fig:rvmeas_results} shows that we can accurately measure RVs for a single star and that the RV bias remains small for binaries with a low mass ratio, i.e. $q < 0.6$, for which the composite spectra are dominated by the lines of the primary. For mass ratios closer to unity, that is when the two stars have nearly equal masses, both stars contribute significantly to the composite spectrum (the one with a blue-shifted profile; the other with a red-shifted profile). We also inspect the simulated spectra visually to investigate if they would be classified as an SB2. Given the resolution and S/N of our MUSE data, the SB2 lines are only sufficiently deblended for RV separations above 200 to 300\,\kms, (depending on the mass and luminosity ratio). At lower RV separations, the measured RVs are strongly biased, resulting in smaller measured RV variations than reality.

Taking these simulations and the difficulties to detect near equal-mass binaries into account, we repeat the bias correction computations. For this, we adopt the measured RVs from the simulated spectra experiment described above in the \vsini=200~\kms\ case as this is representative of most of our stars (see a subsequent paper in this series).
We smooth and interpolate it on a fine grid ranging from 0 to 600~\kms\ in steps of 1~\kms\ in RV separation and from 0.5 to 1.0 by step of 0.01 in mass ratio. This is then used as a correction to the simulations described in Sect.~\ref{s:sb1-bias}, for which the computed primary RV shift in the simulated time series is reduced as a function of the separation between the two components and their mass ratio. Systems with $q<0.5$ are assumed to be unaffected by this bias. Fig.\,\ref{Fig:bias_corr} displays our result and shows that the overall detection probability is reduced by about 10 to 20\%\ compared to the SB1-case. This reduction remains small up to a  mass-ratio of 0.8 but dramatically impacts systems with mass-ratio in the range 0.8 to 1.0 where we can only detect 10 to 20\%\ of the near-equal mass binary population (compared to a detection probability of about 60\%\ without the SB2 bias).

The SB2-bias is particularly dramatic for our present MUSE data. This results from a combination of factors, including the modest spectral resolution, wavelength binning size and S/N of the spectra, the typical (projected) rotation rates, the exponent of the mass-luminosity relation in the B-star regime, and the involved masses and corresponding orbital velocities. Concerning the latter, the reader should be reminded that the orbital velocity of a 7\,\Msun\,+\,7\,\Msun\, binary with a 2-day period is about 200~\kms. Temporal sampling, inclination and longer periods will inevitably push similar systems in the range of velocity separations that are strongly impacted by the SB2-bias. Albeit it would be interesting to investigate the SB2 bias for previously performed O- and B-type binary surveys, we remark that the situation is more favourable for most of the surveys discussed in Sect.~\ref{Sec:intro} and \ref{Sec:discussion}, either because of the higher resolution, S/N, or mass range considered, so that there is no reason to believe that these results are as dramatically affected by the SB2 bias as in the present case.

This is also reflected by the number of detected SB2 systems in this work compared to similar spectroscopic studies of O-type stars. In this study we find only eight out of 282 systems to be SB2s. This is in stark contrast to almost 45\% of SB2s found in a sample of galactic O star \citep{Sana2012}. The number of SB2s detected in \citet{Dunstall2015} and \citet{Banyard2021}, who focus on B-type stars, is similar to what we find here, implying that the mass-luminosity relation in the B-star regime plays an important role in detecting SB2s.

\subsection{Parent orbital parameter distributions} \label{s:pqe_distr}
Integrating the detection probability for flat period and mass-ratio distributions over the range of parameters shown in Fig.\,\ref{Fig:bias_corr} yields an overall detection probability of about 35\%\ when taking into account both the SB1 and SB2 bias discussed in the previous sections.
Adopting different parameter distributions would, however, result in different values. To assess the impact of these choices, we recompute the overall detection probability for different input distributions. In this exercise, we adopt power law representations of the orbital parameters distributions:
\begin{eqnarray}
    f(\log_{10}P/d) &\sim& (\log_{10}P)^{\pi}, \\
    f(q) &\sim& q^{\kappa}, \\
    f(e) &\sim& e^{\eta}
\end{eqnarray}
and we vary their indexes $\pi$, $\kappa$ and $\eta$.

The shape of the period distribution is by far the dominant factor impacting the resulting detection probabilities. Fixing all other assumptions, and changing $\pi$ from 0.0 to $-0.5$  modifies the overall detection probability from 0.35 to 0.45 given that the period distribution now favours shorter-period systems that are easier to detect. Varying $\kappa$ and $\eta$ from $-0.5$ to $+0.5$, respectively, has an overall impact of a few 0.001 and is thus negligible in this context.

To adopt meaningful parameter distributions, we consider measurements obtained using the same parameter ranges as the ones we adopt here\footnote{We refer to \citet{Almeida2017} for a discussion on how the boundaries of the orbital parameter intervals impact the measured power-law indexes of the orbital parameter distributions.}.  O-star studies in the Galaxy and LMC yielded $\pi=-0.55\pm0.22$ and $\pi=-0.45\pm0.39$,  respectively \citep{Sana2012,Sana2013}. The LMC  value was revised to  $\pi\approx -0.2$ by \citet{Almeida2017} using a more comprehensive data set. Finally, \citet{Dunstall2015} obtained $\pi=0.0\pm0.5$ for early B-type stars in the LMC. We therefore adopt a value of $\pi=-0.25 \pm0.25$, which encompasses all measured values in its $\pm1\sigma$-interval, as a plausible representative value for the period distribution. Neither \citet{Sana2013} nor \citet{Dunstall2015} could reliably constrain $\kappa$ and $\eta$, so we adopt the values from the Milky Way O-star sample, that is $\kappa=-0.2\pm0.6$ and $\eta=-0.4\pm0.2$ \citep{Sana2012}, but note that the latter choices have no impact on our results.

Given that the period index $\pi$ was proposed to be mass-dependant \citep[i.e., skewed towards shorter periods for O stars, and to slightly longer periods for late-B stars;][]{Rizzuto2013, Moe2017, Tokovinin2020}, we test how such a mass-dependant period index affects our estimate of the detection probability. We therefore assume a variable period index that changes linearly from $\pi=-0.25\pm0.25$ at the upper mass end considered here, to $\pi=+0.25\pm0.25$ at the lower mass end. We find that the obtained detection probability is lower by 0.06 which remains within the error bars of a mass-independent detection probability computed with using $\pi=-0.25 \pm0.25$ for the entire considered mass range (see Table\,\ref{Tab:bin_stats}). Therefore we adopt the latter for the following computations.

In the final set of bias correction computations, we thus draw 10\,000 artificial populations of 280 binaries (corresponding to our sample size). For each population, we randomly assigned values of $\pi, \kappa$ and $\eta$ from the parent orbital distributions following Normal distributions with central values and 1$\sigma$ dispersion taken as described above. The detection probability and, after a second iteration its error bars, are then obtained as the average and root-mean-square of the fraction of detected binaries across the 10\,000 simulated populations. In that sense, the error bars encompass both the uncertainties due to the choice of orbital distributions and due to the sample sizes. By running simulations with much larger population sizes, we estimate that the adopted spread in orbital parameter distributions amounts to about 6\% in the final uncertainty.

Finally, we also computed a magnitude-dependant bias correction for which we pick primary masses corresponding to the different magnitude bins as well as typical RV errors measured for stars in that magnitude bin (see Appendix \ref{App:bias}), as well as in different regions of the CMD (see below). 
Our bias correction does not take into account the evolutionary history of binary systems and the nature of possible companions, but is purely based on the assumed distributions of orbital parameters. Given that the adopted distributions are typical of young massive-star populations, they are more likely to represent the initial conditions than more evolved populations that have been impacted by effects of binary interactions.

\section{Multiplicity properties}\label{Sec:binarity}
The overall observed spectroscopic binary fraction, that is of our entire sample of 282 stars with measured RVs, is $f_\mathrm{SB}^\mathrm{obs} = 13.2 \pm 2.0\%$. Applying the bias correction described in Sect.\,\ref{subsec:obsbias} leads to an intrinsic, bias-corrected close binary fraction of $f_\mathrm{cl} = 34^{+8}_{-7}$\% for the massive-star population of NGC~330.

\begin{figure}\centering
    \includegraphics[width=0.99\hsize]{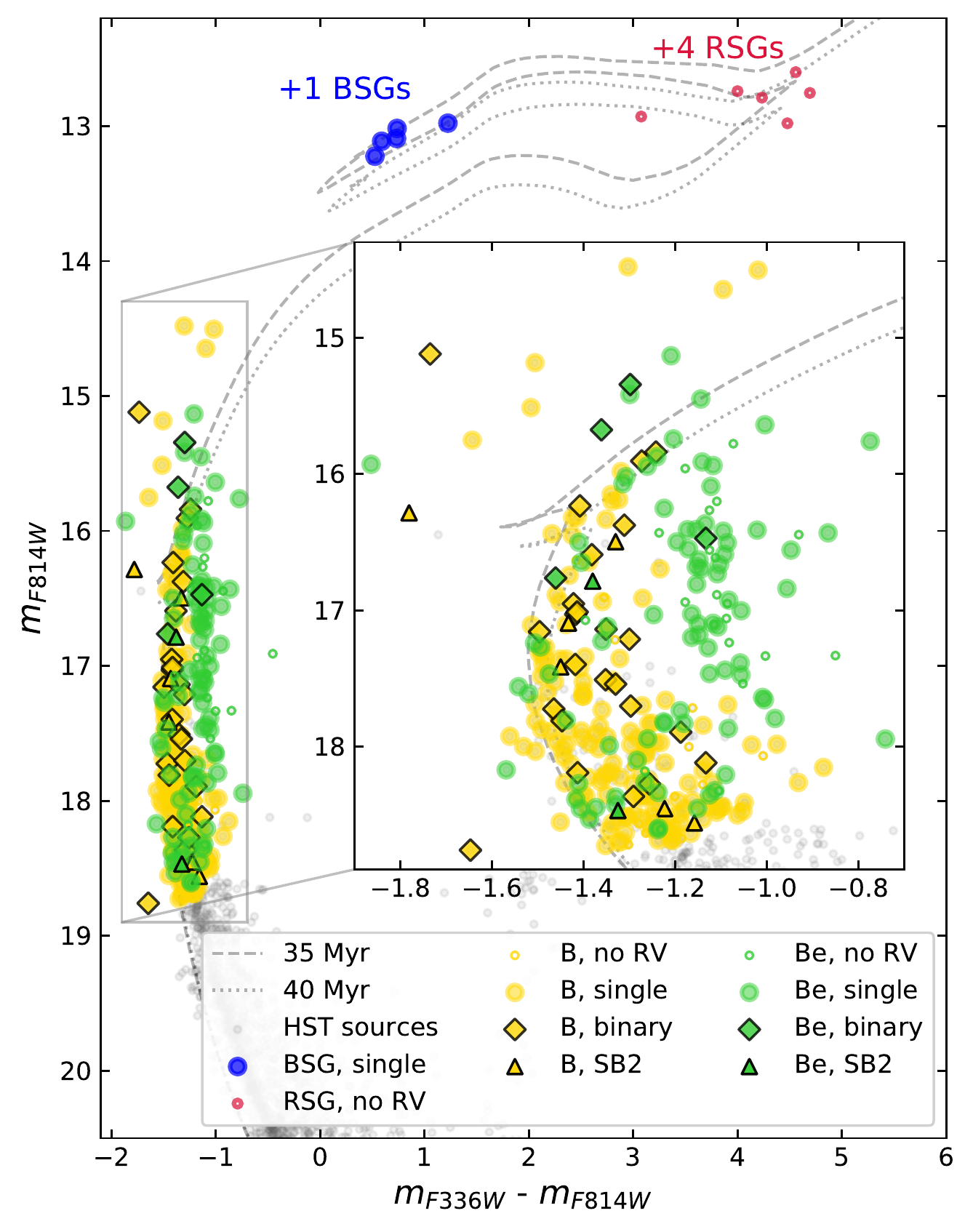}
    \caption{\label{Fig:CMD_bin} HST color-magnitude diagram  \citep{Milone2018} overlaid with binarity information. B-type stars are marked in yellow, Be stars in green, BSGs in blue, and RSGs in red. Open small circles indicate that no RVs could be measured, filled circles are stars with no significant RV variability, diamonds indicate stars classified as binary systems based on their RVs, and triangles indicate SB2 candidates.}
\end{figure}

Combing the spectroscopic information from MUSE with the photometric catalogue of \citet{Milone2018}, Fig.\,\ref{Fig:CMD_bin} displays a color-magnitude diagram (CMD) with binarity information. It shows that most of the spectroscopic binaries are situated on the MS. There are, however, also a handful of binaries among the Be stars, and several binaries are located above the MS turnoff, as indicated by the single-star non-rotating Padova isochrones \citep{Bressan2012, Chen2014, Chen2015, Tang2014, Marigo2017, Pastorelli2019}. There are no magnitudes in the relevant HST filters for the binary BSG, nor for the four additional RSGs, so they are not included in the CMD.

Sect.\,\ref{Subsec:bin_tot} describes our results for different spectral (sub)types. In Sect.~\ref{Subsec:bin_cmd}, we use the position of stars in the CMD to discuss the multiplicity of different evolutionary stages. A more detailed description of several interesting binary systems is given in Sect.\,\ref{Subsec:objects}.

\subsection{Binary fraction of the massive-star population}\label{Subsec:bin_tot}


Below, we briefly discuss the binary fraction of the B-type stars, Be stars and BSGs in our sample. 
In this context, we refer to Be stars as all stars that show Balmer line emission in their spectra (see also Sect.\,\ref{Subsec:bin_cmd}).

\textit{B stars:} \quad Among the 209 B stars in our sample, 165 stars do not show significant RV variations, while 24 stars are classified as binaries based on their RV variation. Among the 165 non-RV variable stars, 6 show composite line profiles indicative of SB2 systems. RV measurements were not possible for 20 stars. Combining the RV variables with the 6 detected B-type SB2s leads to an observed spectroscopic binary fraction of $f_\mathrm{SB, B}^\mathrm{obs} = 15.9 \pm 2.6$\%.

\textit{Be stars:} \quad 93 of the 115 Be stars in our sample could be investigated with RV measurements. Only five Be stars pass our RV-variability criteria and are considered as binaries. Additionally, we classify 2 Be stars as candidate SB2s. Putting this all together, we find an overall observed spectroscopic binary fraction of $f_\mathrm{SB, Be}^\mathrm{obs} = 7.5 \pm 2.7$\% for the Be stars in the sample.

\textit{Blue supergiants:} \quad One of the six BSGs in the sample, namely \#\,729 shows significant RV variations and is classified as binary following our multiplicity criteria. This yields an observed spectroscopic binary fraction of BSGs in the core of NGC~330 of $f_\mathrm{SB, BSG}^\mathrm{obs} = 17 \pm 15\%$. The large errors result from the low number statistics. 

\begin{figure}[t!]\centering
    \includegraphics[width=\hsize]{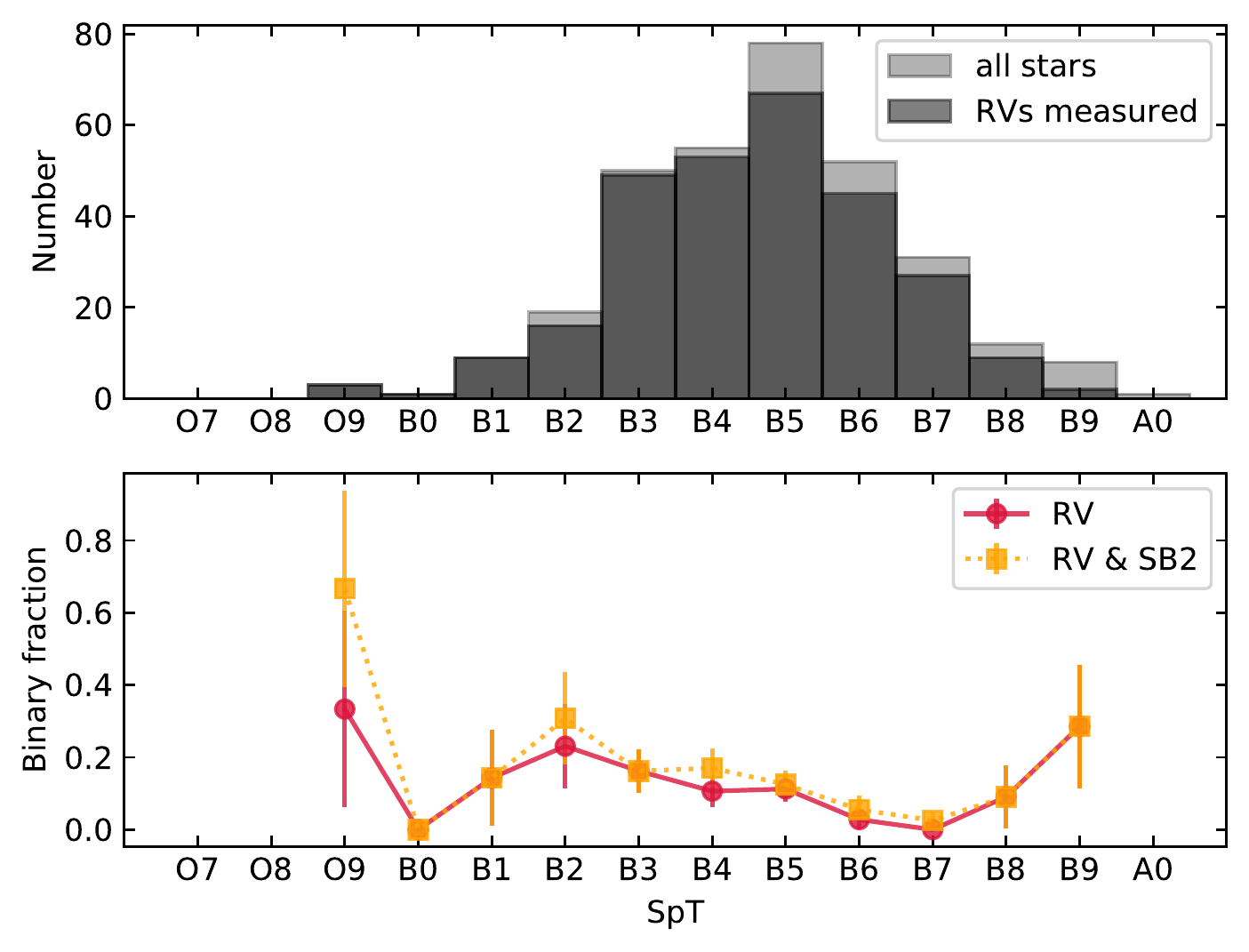}
    \caption{\label{Fig:binfrac_spt} Top panel: distribution of spectral types for all main sequence stars (dim gray) and stars for which RVs were measured (dark gray). Bottom panel: observed spectroscopic binary fraction as a function of spectral type, where the red curves indicates the binary fraction inferred from the RV measurements while the orange curve combines the RV measurements with the SB2s found by visual inspection.}
\end{figure}

Figure\,\ref{Fig:binfrac_spt} shows the distribution of spectral types and the observed close binary fraction for B and Be stars. Spectral sub-type bins with less than 10 objects provide few statistical constraints (i.e., spectral types earlier than B2 and later than B8) and will not be further considered. The remaining spectral sequence, from B2 to B7, shows a clear decrease in the observed binary fraction. This trend is more obvious when including the SB2s, but can also be seen from the RV measurements alone. When including the SB2s, the observed binary fraction decreases from 30\% at B2 to less than 3\% at B7. Given the limitations of our data, it is hard to decide if the increase in the binary fraction towards the latest spectral types, that is B8 and B9, is real or due to the low number statistics. 

As discussed in \citetalias{Bodensteiner2020a}, the spectral classification of the Be stars in the sample is difficult. On the one hand, the \ion{He}{i} lines used for spectral classification can be influenced by emission infilling from the Be star disk. On the other hand, Be stars are rapidly rotating stars which, depending on the inclination angle, may appear cooler due to their rotation. This might lead to a classification into later spectral types, which impacts the spectral type distribution. We therefore also consider the observed spectroscopic binary fraction as a function of F814W magnitude, which is less sensitive to these effects.

\begin{figure}[t!]\centering
    \includegraphics[width=\hsize]{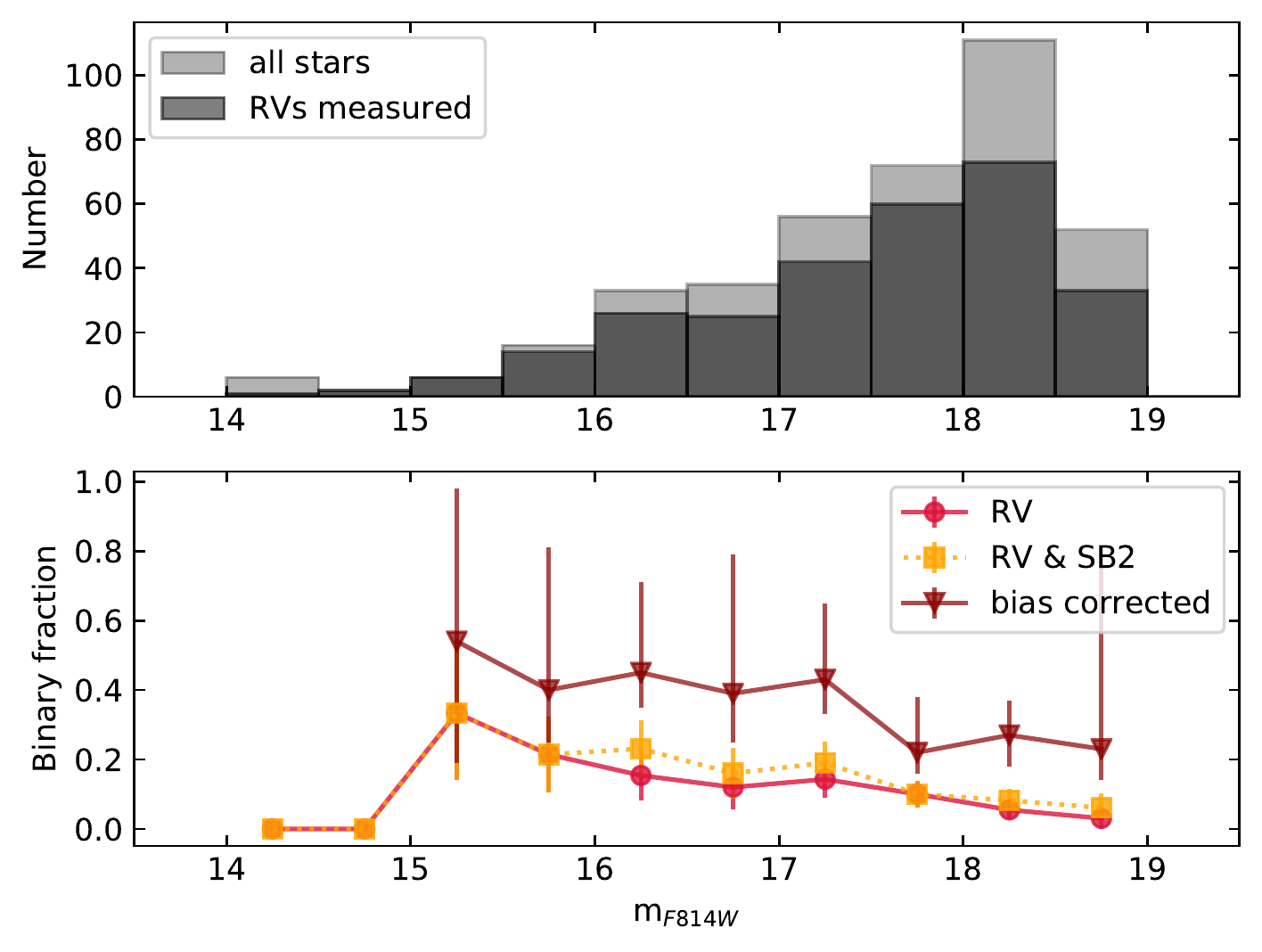}
    \caption{\label{Fig:binfrac_mag} Top panel: distribution of F814W magnitudes of all stars with for which spectra were extracted (dim gray), and stars for which RVs were measured (dark gray). Bottom panel: observed spectroscopic binary fraction as a function of F814W magnitude where the orange curve is based on RV measurements and the red on RV measurements as well as SB2s. The bias-corrected close binary fraction is overplotted in dark red.}
\end{figure}

Figure\,\ref{Fig:binfrac_mag} shows the distribution of F814W magnitudes of the B and Be stars as well as their binary fraction. The distribution of F814W magnitudes indicates that our sample is not complete at F814W>18.5\,mag. As shown in the bottom panel, the decreasing trend in the observed binary fraction that was visible in the spectral type distribution is also visible here. Neglecting stars brighter than 15$^{\rm th}$ magnitude where the number statistics are low, as well as stars fainter than F814W = 18.5\,mag where the sample is incomplete, we find that the observed spectroscopic binary fraction $f_{SB}^{obs}$ decreases from 33\% to 7\%. 

We apply a magnitude-dependant bias correction to the observed binary fraction, as described in Sect.\,\ref{s:pqe_distr}. We find that the bias-corrected binary fraction of the brightest stars with F814W=15\,mag, which corresponds to about 8.5\Msun\, is almost 60\%. It drops down to about 25\% for the faintest stars considered here, which roughly corresponds to 5\,\Msun.
While uncertainties in each magnitude bin are large due to smaller sample sizes, the downward trend remains, hence it cannot be explained by the observational biases alone. At least part of it is likely due to genuine differences in the multiplicity properties. This decrease in the close binary fraction can either be due to a genuine lack of binaries at later spectral type, or by a shift in the period distribution towards larger orbital periods.

In order to further investigate whether this decreasing trend in the bias-corrected close binary fraction is statistically significant, we perform a linear regression between F814W=15\,mag and F814W=18.5\,mag, and find that we can reject the null-hypothesis that there is no trend in the data at the 5\%-level. Our data are insufficient to decide whether the decrease of the close binary fraction with mass is linear or as a different functional form. A visual inspection of Fig.\,\ref{Fig:binfrac_mag} may indeed suggest a step at F814W $\simeq 17.5$, corresponding to $\simeq 6.5$\,\Msun on the MS.

\subsection{Binarity in different regions of the CMD}\label{Subsec:bin_cmd}

In order to compare the observed close binary fraction in different evolutionary stages, we investigate the binary properties of the massive-star population from a different angle, that is by defining different regions in the CMD based on expectations from theoretical considerations.

\begin{figure}[t!]\centering
 \includegraphics[width=\hsize]{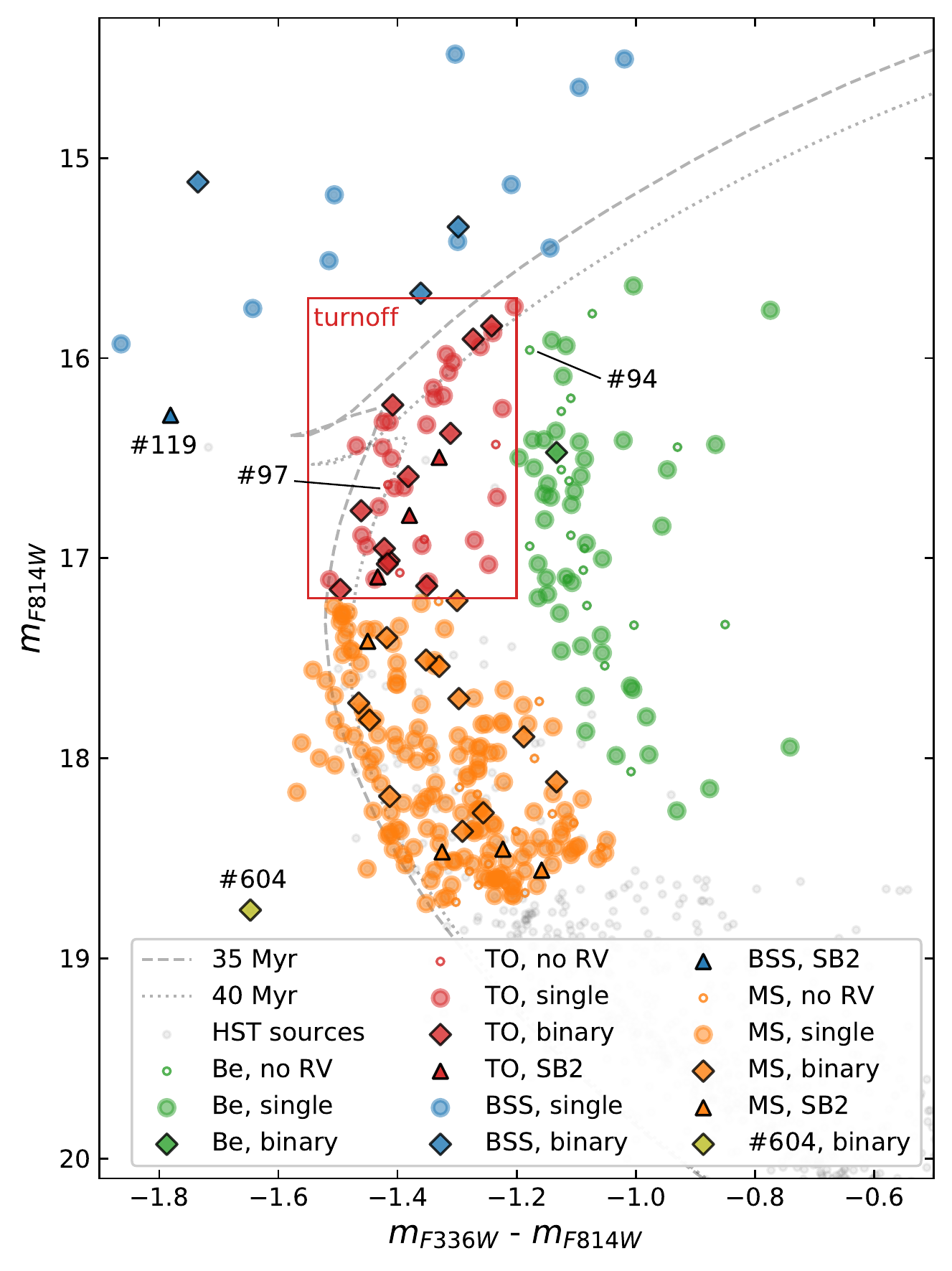}
    \caption{\label{Fig:cmd_regions} Binarity of different stellar groups, based on their position in the CMD. MS stars are marked in red, stars in the Be star region in dark green, stars around the cluster TO in red, and blue stragglers in blue. Additional HST sources for which no spectra were extracted are marked with gray dots, and stars without RV measurements are indicated with open circles. Colored filled circles indicate that the star is single, colored diamonds indicate the star is classified as binary based on RV variability, and colored triangles indicate SB2s. The individual systems of interest discussed in Sect.\,\ref{Subsec:objects} are labeled.}
\end{figure}

Apart from the BSGs and RSGs, we use the Padova isochrone at 40\,Myr to define four additional regions in the CMD, as illustrated in Fig.\,\ref{Fig:cmd_regions}:
\begin{enumerate}
    \item the MS region, which is an extended band with a width of about 0.4 magnitudes around the 40\,Myr isochrone;
    \item the Be star region, which appears as a second group, red-wards of the MS, by a color-offset of about 0.3 magnitudes in Fig.~\ref{Fig:cmd_regions};
    \item the turnoff (TO) region, which is located around the cluster turnoff of the 40\,Myr Padova isochrone;
    \item the blue straggler star (BSS) region, which is above and to the left of the cluster turnoff.
\end{enumerate}
While the BSS and the Be star region are clearly identifiable from the CMD, the definition of the TO region is more subjective. Furthermore, the assignment of stars to different groups is purely based on the position in the CMD. This means that, for example, a star observationally classified as a Be star, situated in the blue straggler region, will be assigned to the blue straggler region. Similarly, we classify all stars as Be stars that are in the Be star region, even if they do not show emission in our spectra (this is the case for only five targets at the faint end of the Be star region). There is one star, classified as binary based on RV variability that is to the left of the MS. This source will be discussed in Sec.\,\ref{Subsec:objects}.

Table\,\ref{Tab:bin_stats} presents the spectroscopic and bias-corrected close binary fraction for each of these groups separately, including binaries detected through RV variations or through their SB2 line profiles. We also list the binary properties of the BSGs (see Sect.\,\ref{Subsec:bin_tot}). The RSGs were studied by \citet{Patrick2020}, who probed a different, complementary parameters space than ours. Apart from improving on the time coverage, 
they further use multi-epoch ESO-HARPS spectra (with a resolving power of R$\sim$80,000).  Additionally, their study includes five RSGs that are outside of the MUSE FoV.

For each of the regions in the CMD, we compute the binary detection probability, taking into account the approximate masses of the stars populating different regions as well as typical RV errors as a function of F814W magnitude and color. The difficulty of measuring RVs for Be stars, whose spectra are dominated by emission, becomes apparent here: while populating a similar brightness range as TO and MS stars, the detection probability is significantly lower for them, especially compared to the brighter TO stars.

In this exercise, no attempt was made to account for the impact of stellar evolution on the distribution of orbital parameters, i.e. all simulations were performed adopting the orbital parameter distributions derived from young populations (see Sect.\,\ref{s:pqe_distr}). Populations most impacted by binary interaction, that is Be stars and BSSs, are likely to deviate from these assumptions \citep[e.g.,][Wang et al., in prep.]{Langer2020}.

Table~\ref{Tab:bin_stats} provides the bias-corrected close binary fraction for the entire sample (which includes the before-mentioned caveat) as well as for stars in the MS and TO region. For the latter, we provide the detection probability computed with our standard computation assuming a constant index $\pi$, as well as assuming a mass-dependent $\pi$ as discussed in Sect.\,\ref{s:pqe_distr}. For the Be stars and BSSs, we only provide the detection probability and not the bias-corrected binary fraction, so that the results presented in Table~\ref{Tab:bin_stats} still provide a valuable relative comparison of the impact of variable RV measurement accuracy for different categories of stars given their respective observational specificity discussed in Sects.~\ref{Sec:ObsAndSpec} and \ref{Sec:RVmeas}. The different spectroscopic and intrinsic close binary fractions of the different groups will be discussed in Sect.~\ref{Sec:discussion}.

\renewcommand{\arraystretch}{1.2}
\begin{table}\centering
    \caption{\label{Tab:bin_stats} Binary fractions in different regions of the CMD. The first column indicates the region, the second column the number of stars that are assigned to the region, and the third column the observed binary fraction $f_\mathrm{SB}^\mathrm{obs}$. The last two columns give the detection probability p$_\mathrm{detect}$ and bias-corrected binary fraction $f_\mathrm{cl}$. "ALL" refers to the all stars studied in this work, that is MS + TO + Be + BSS stars as well as the BSGs.}
    \begin{tabular}{lrrlr} 
        \hline \hline
        region & N & $f_\mathrm{SB}^\mathrm{obs}$ [\%] & p$_\mathrm{detect}$\tablefootmark{\#} & $f_\mathrm{cl}\tablefootmark{\#}$ [\%]  \\ \hline
        ALL   & 284 &  $13.2\pm2.0$ &  $0.39\pm0.05$ & $34^{+8}_{-7}$\\
        MS+TO & 217 &  $13.8\pm2.3$ &  $0.40\pm0.08$ & $35^{+10}_{-6}$\\
              &     &               &  $0.34\pm0.06$\tablefootmark{$\ddagger$} & $40^{+9}_{-6}$\\
        \hline
        MS & 175 & $9.1 \pm 2.2$ & $0.38\pm0.06$ & $24^{+10}_{-6}$ \\
        TO & 42 & $33.3 \pm 7.3$ & $0.57\pm0.11$ & $59^{+20}_{-15}$\\
        Be & 47 &  $2.1 \pm 2.1$ & $0.29\pm0.08$ & \tablefootmark{\textdagger} \\ 
        BSS & 14 & $28.6 \pm 12.1$ & $0.57\pm0.15$ & \tablefootmark{\textdagger} \\ 
        BSG & 6 & $17 \pm 17$ &  $0.61\pm0.20$ & $28^{...}_{...}$ \\
        RSG\tablefootmark{*} & 9 & - & - & $30 \pm 10$ \\
        \hline
    \end{tabular}
    \tablefoot{
    \tablefoottext{\#}{These detection probabilities and bias-corrected binary fractions are valid under the adopted parent orbital parameter distributions discussed in Sect.\,\ref{s:pqe_distr}. The quoted 1$\sigma$ errors contain both the statistical uncertainties due to the sample size and those due to random drawing of the index of the parent distributions.}\\
    \tablefoottext{\textdagger}{While providing the detection probability also for Be stars and BSS, we do not propagate it further to compute the intrinsic binary fraction of these groups of stars as the fraction of BiPs among them is probably significant.\\}
    \tablefoottext{$\ddagger$}{{\bf Detection probability computed assuming a linearly variable period index from $\pi=-0.25 \pm .25$ at the upper mass range to $+0.25 \pm .25$ at the lower mass range.}}\\
    \tablefoottext{*}{From \citet{Patrick2020}, including both RSGs in the core and in the outskirts, based on higher-quality ESO-HARPS data with longer time coverage.}}
\end{table}

\subsection{Individual systems of interest}\label{Subsec:objects}
\paragraph{The binary BSG} \#729 or Cl*~NGC~330~ARP~35 was first studied by \citet{Arp1959} and classified as B9I by \citet{Feast1964}. In \citetalias{Bodensteiner2020a}, we classify it as early A Ia-Ib. \citet{Feast1980} measured RVs and found values between 134 and 154\,\kms, which are of a similar order of magnitude than the RVs we measure (i.e., a mean value of 152.6 \kms). Based on their data, they do not classify it as (candidate) binary. Using the MUSE observations, we find that \#729 is clearly RV variable and according to our two criteria, we classify it as binary. Given that we only have 6 epochs, it is not possible to derive orbital parameters such as the period of the system. We see no contribution of the secondary in the spectrum. Given that \#729 was not included in the HST input catalogue (probably because of saturation) we can not add it to the CMD in Fig. \,\ref{Fig:CMD_bin}.

\paragraph{A Be X-ray binary candidate} In \citetalias{Bodensteiner2020a}, we reported on the detection of a high-mass X-ray binary detected in the MUSE FoV \citep{Shtykovskiy2005, Sturm2013, Haberl2016}. Using the multi-epoch data, we investigate the binary status of the two stars (i.e. \#94 and \#97, both labelled in Fig.\,\ref{Fig:cmd_regions}) that are within the 0.7" error on the position of the X-ray source. Both are classified as Be stars. While \#97 shows no significant RV variations, we cannot measure RVs following the procedure described in Sect.\,\ref{Sec:RVmeas} for \#94 given that all available \ion{He}{i} absorption lines used for RV measurements are filled-in with emission. A preliminary RV estimate, using the Balmer emission lines in \#94, however, shows significant RV variations. Based on the six epochs of MUSE observations, it is unfortunately neither possible to derive an orbit, nor to prove association of either of the two Be stars with the X-ray source. Follow-up observations will further investigate this possible Be X-ray binary.

\paragraph{An O+B binary} One of the few O stars in the sample, \#119, classified as such because of the presence of \ion{He}{ii} absorption, is also an SB2. The spectra of \#119 are composite and show line-profile variations indicative of a B star companion. Additionally, the \halpha\,line shows significant infilling, while the H$\beta$ line shows only absorption. Given that there is a \halpha-emitting source close-by, this could be contamination. It could also imply the presence of a disk, either around one of the two stars, or around the binary system. According to the position in the CMD (see Fig. \,\ref{Fig:cmd_regions}), the system is in the blue straggler region. This region is populated by objects thought to be rejuvenated in binary interactions, and currently interacting systems \citep{Wang2020}. Given that we detect it as a close binary system, and that in the case of ongoing mass transfer material may be lost from the system, it is conceivable that \#119 has a circumbinary disk. More observations are, however, required to further investigate the nature of this system.

\paragraph{A binary system hotter than the MS} \#604, marked by a yellow diamond in Fig.\,\ref{Fig:cmd_regions}, stands out in the CMD. Being to the left of the MS implies that it is significantly hotter than stars of similar brightness. The spectral type we determine from the MUSE data is approximately B5, implying that the star is not particularly hot. This is, however, based on a rather low S/N (between 60 and 85 in the different epochs), as the star is relatively faint (F814W = 18.8 mag). Given that the star is close to the edge of the MUSE FoV, spectra could only be extracted in five of the six epochs. The average RV measured for this star is $58\pm8$\kms, and thus differs significantly from the mean RV of the SMC of $153.7 \pm 1$\kms\,\citep{Patrick2020}, that is by almost 100 \kms. The position in the CMD as well as the RVs differing from the RV of the SMC may imply that \#604 is a foreground star. Such magnitudes are, however, also indicative of a hot, envelope-stripped companion \citep[e.g.,][]{Gotberg2018}. 
We classify \#604 as a binary system based on the measured RV variations. In the MUSE spectra, we see no indication of a secondary, and further observations with higher resolution and better S/N ratio will be necessary to test this hypothesis.

\section{Discussion}\label{Sec:discussion}

\subsection{Different binary fractions in different regions of the CMD}
As summarized in Table\,\ref{Tab:bin_stats}, we find that the spectroscopic binary fraction varies strongly in different regions of the CMD, that is as a function of stellar mass and evolutionary status. While we refrain from computing the bias-corrected binary fraction of Be stars and BSS, the bias-corrected binary fraction, for a given set of parent orbital parameter distributions, is about 25\,\% on the lower MS and increases to approximately 60\% around the cluster turnoff. In the more evolved evolutionary phases, the BSGs and the RSGs, the binary fraction is 28\% and 30\%, respectively. However, as noted above, the orbital periods for RSGs probed by \citet{Patrick2020} is quite different. These changes are statistically significant and indicate a change in the binary fraction of the population, or in its distribution of orbital periods.

While it is difficult to disentangle the effects of initial mass and evolutionary status, we compare our findings to expectations from binary evolution theory \citep[see e.g.,][]{Wang2020}. Most binary systems below the cluster TO are expected to be pre-interaction systems. In the hot part of the HRD, binary interactions occur preferentially around the TO, that is when the stars expand more significantly toward the end of their MS lifetime. Around the TO, the number of currently interacting systems is expected to be high, while the BSS region is populated entirely by post-interaction systems. As discussed in the introduction, Be stars are often interpreted as binary interaction products \citep{Pols1991, Shao2014, Bodensteiner2020b}, implying that their companions are expected to be either envelope-stripped stars \citep{AbdulMasih2020, Shenar2020, Bodensteiner2020c, ElBadry2020a, ElBadry2021} or compact objects \citep{Langer2020,Schurman2021}. Both types of objects are difficult to detect spectroscopically and most BiPs would not be captured in our bias correction due to the nature of the companions and the expected long orbital periods (Sect.\,\ref{subsec:obsbias}).

\textit{MS region:} \quad The vast majority of stars on the MS are thought to either be single stars, or binary stars that likely have not interacted yet. Their companions are most likely also MS stars, and we find a bias-corrected close binary fraction of $f_{cl}\,=\,24^{+10}_{-6}$\%. On higher resolution data, depending on their mass ratio and S/N, such MS companions would be relatively easy to detect. Moreover, we have shown in Sect.~\ref{s:sb2-bias} that our MUSE data suffer from significant detection bias against near-equal mass SB2 systems. This may explain the lack of detected binaries close to the red edge of the extended MS \citep[the so-called binary line,][]{Milone2018}, where binary systems with a mass ratio close to unity are expected.

\textit{TO region:} \quad Around the cluster TO, the observed binary fraction reaches $33\pm7$\%\ and $59^{+20}_{-15}$\% after bias-correction. The detection capability of our campaign is better for stars in the TO region than for MS stars due to their higher S/N and higher masses. 
However, this is insufficient to explain the large difference compared to MS stars, that is over a factor of two, even after bias-correction (see Table~\ref{Tab:bin_stats}). As described above, binary interactions occur preferentially toward the end of the MS lifetime of stars which would allow them to remain around the TO longer than if they were single stars.
The region around the TO is thus probably mainly populated by two groups of binaries: pre-interaction binaries (e.g., nearly-equal mass binary systems which appear brighter than single MS stars), as well as currently interacting systems (e.g., slow case-A binaries).
Additionally, there might be a difference of intrinsic multiplicity properties between the TO and MS regions owing to their different mass ranges  ($\sim7$\Msun\,$\it vs.~\sim5$\Msun). On the one hand, the binary fraction increases with spectral types \citep[e.g.,][]{Moe2017}. On the other hand, solar-like stars typically have longer orbital periods than massive OB stars. Therefore, somewhere in between the two mass regimes, a transition should occur that will mostly impact the period distribution, hence the detection probability of our survey. 

\textit{Be star region:} \quad The low observed spectroscopic fraction of binaries in the Be star region, which is only 2\%, could have two explanations: either the close binary fraction of classical Be stars is intrinsically low, or Be binaries have long orbital periods, for which our detection probability drops (see Fig.\,\ref{Fig:bias_corr}). These findings are compatible with qualitative expectation of the binary scenario.  Of course, our RV study cannot distinguish post-interaction Be products with genuine single Be stars. Yet, our simulations show that, should Be stars have a MS companion with a mass and period distribution similar to the MS B-type star population, then we would have detected a significant fraction of Be binaries. In that sense, our results are also compatible with the conclusions of \citet{Bodensteiner2020b}, that there is a dearth of Be+MS binaries.

\textit{BSS region:} \quad Stars brighter and hotter than the cluster TO, i.e. blue stragglers, are typically interpreted as binary interaction products. They are either merger products, stars that were rejuvenated as mass gainers, or currently interacting systems. While merger products are now single stars (disregarding the possibility of initial triple systems), currently interacting (Algol) systems have a large probability to be flagged as binaries in our study. As mentioned before, given that the BSS region is populated by BiPs, we do not apply a bias correction here. We do, however, find that the observed spectroscopic binary fraction of the BSS and TO populations are similar in our study.

\subsection{Comparison to the outskirts of NGC~330}
Our MUSE observations target the inner $1'\times1'$ core of NGC~330. The OB star population in the outskirts was investigated with VLT/FLAMES in \cite{Evans2006} down to $V=17$~mag, reporting an observed spectroscopic binary faction of only 4\%, i.e. considerably smaller than our overall fraction of 13\%\ in the cluster core. As described in \citetalias{Bodensteiner2020a}, the overlap between the two samples in terms of brightness is limited. More specifically, there are only 27 stars in our MUSE sample that are brighter than $V=17$\,mag, indicating that our comparison may suffer from low number statistics.

When restricting our sample to the 27 stars brighter than V=17\,mag, we find that the observed spectroscopic binary fraction in the cluster core is $f_\mathrm{SB, V<17}^\mathrm{obs} = 18.5 \pm 7.5$\%. This is in strong contrast to the 4\% reported by \citet{Evans2006} for the outskirts. This might, however, be due to the relatively short time coverage of the spectra from \citet{Evans2006} of only $\sim$10 days, which means that their study is only sensitive to the shortest orbital periods.

\subsection{Comparison to other clusters}
Comparing the close binary fraction in NGC~330 to the one measured in other clusters allows us to investigate the effects of two important additional cluster properties: the cluster age and the environment metallicity. Indeed, it remains unclear what their respective impact is on the multiplicity properties. As described in the introduction, previous multiplicity studies have mainly focused on O- and early B-type stars in the MW and LMC environments. These OB stars are, of course, only present in young clusters. As discussed in Sect.\,\ref{s:sb2-bias}, the SB2 bias investigated here has a less significant impact on the results of these works given the higher resolution of the data used.

\citet{Kobulnicky2014} investigated the binary properties of massive stars in the 3-4\,Myr old Cygnus OB2 association in the Milky Way. Focusing on a sample of around 130 O and early-B type stars (spectral type earlier than B2) they observe a close binary fraction of $\sim$35\%. Correcting for observational biases, they find an intrinsic binary fraction close to 55\%, which is higher than what we find for NGC~330. This is probably due to the fact that the stars included in their sample are significantly more massive, and the binary fraction increases as the mass of the primary increases \citep[see e.g.,][and Fig.\,\ref{Fig:binfrac_mag}]{Moe2017}.

\citet{Banyard2021} studied the B-star population of the young, galactic cluster NGC~6231, which has an estimated age between 4 and 7\,Myrs. They report an observed spectroscopic binary fraction of $35 \pm 4$\%, which, corrected for observational biases using the same method as described here leads to an intrinsic close binary fraction of $44 \pm 6$\%. This is slightly higher than the bias-corrected binary fraction we measure in NGC~330, also when only considering MS+TO stars. There are two possible reasons for this: first, the population of stars studied by \citet{Banyard2021} is more massive. Alternatively, it might be due to the age difference of the two clusters. While a young cluster like NGC~6231 allows us to probe close to the initial binary properties, NGC~330 is at an age where a significant number of binary systems had time to interact. As described above, our bias correction focuses on MS companions and does not correct for post-interaction systems, which are extremely difficult to detect from the observations.

\citet{Dunstall2015} measured the binary fraction of B stars in the 30 Doradus region in the LMC. Focusing here on the non-supergiant population, they report an observed spectroscopic binary fraction of $25\pm2$\%. Additionally, they provide measurements for individual clusters of different ages \citep{Evans2015}: they observe spectroscopic binary fractions of $27\pm5$\% and $34\pm5$\% for the younger clusters NGC~2070 and NGC~2060 \citep[age $\simeq 2-8$\,Myr,][]{Schneider2018}, respectively. The observed spectroscopic binary fractions reported for the slightly older clusters Hodge~301 and SL~639, with estimated ages of 15$\pm$5\, Myr and 10-15\,Myr \citep{Evans2015}, are only $_f\mathrm{SB}^\mathrm{obs} = 8\pm8$\% and $10\pm9$\%. They estimate the binary detection probability of B-type stars to be 0.4. Correcting the observed binary fraction for the detection probability leads to an intrinsic close binary fraction of
$58\pm11$\% for the overall B-star population, and an intrinsic close binary fraction of $20.0\pm20$\% and $25.0\pm22.5$\% in Hodge~301 and SL~639, respectively. Despite the low number statistics in Hodge~301 and SL~639 (i.e., 22 stars in total), the intrinsic binary fractions are very similar to the bias-corrected binary fraction of MS stars reported here (i.e. $24^{+10}_{-6}\%$).

As mentioned before, both the older age of NGC~330 as well as the lower metallicity of [Fe/H]\,$\lesssim\,-1.0$ could be the reason for the lower inferred binary fraction in NGC~330. Several previous studies have found that, for massive stars, the close binary fraction as well as the orbital parameter distribution seem universal over different metallicities \citep[see e.g.,][]{Sana2013, Moe2013, Sana2017, Almeida2017, Banyard2021}. Given that we cannot independently measure the orbital parameter distributions of the B-type stars in NGC~330 with the data discussed in this work, our bias-corrected close binary fractions are based on the assumption of those. The present data set does therefore not allow for a clear distinction between the two scenarios.

Future studies comparing the results of NGC~330 to clusters of similar age in different metallicity environments would allow to investigate any possible metallicity effect. Furthermore, determining the orbital solutions of the detected binary systems in NGC~330 would allow to map the orbital parameter distributions of B-type stars at SMC metallicity. These would allow to investigate whether the lower binary fraction in NGC~330 is indeed an age effect or due to the lower metallicity. The binary fraction of MS+TO stars obtained here are, however, in general agreement with the binary fractions found for Galactic and LMC samples.

\section{Summary and future work}\label{Sec:conclusion}
In this second paper of a series, we use multi-epoch MUSE observations in wide-field mode and supported by AO to investigate the multiplicity properties of the massive-star population of the $\sim40$~Myr-old, massive, open cluster NGC~330 in the SMC. We extract spectra for 400 stars and measure RVs of almost 300 stars using Gaussian fitting of spectral lines. Based on two RV variability criteria, we classify stars as single or binary stars. We furthermore visually inspect the spectra of all stars in order to detect composite spectra and line profile variations indicative of SB2s. We therefore provide the first large and homogeneous RV study of B-type stars in a metal-poor environment.

We find that the overall observed spectroscopic binary fraction of the massive-star population in NGC~330, including B stars, Be stars, and BSGs, is $f_\mathrm{SB}^\mathrm{obs} = 13.2 \pm 2.0\%$. To correct for observational biases, we simulate two effects: the  overall binary detection sensitivity of our observing campaign based on the amplitude of the primary RV motion (called SB1 bias), as well as the difficulty to detect binary systems with similar components (referred to as SB2 bias). Given mainly the low-resolution of the MUSE spectra and the masses of the stars considered here, we find that the SB2 bias is sizeable in this study, and probably much more important than in previous studies that have implemented the SB1-bias only \citep[e.g.,][]{Sana2012, Sana2013, Dunstall2015, Banyard2021}.

Correcting for these observational biases, we find a bias-corrected close binary fraction of $f_\mathrm{cl} = 34^{+8}_{-7}$\% for the massive-star population in NGC~330. Considering the B and Be stars, we find that the observed as well as bias-corrected close binary fraction is a function of F814W magnitude: the bias-corrected close binary fraction drops from $\sim$55\% to $\sim$25\% in the considered magnitude range. While the brightest stars are above the cluster turnoff, and thus most strongly impacted by binary interaction, this might provide indications that the binary fraction drops as a function of mass. Disentangling these two effects is, however, not possible based on the current data.

Investigating the binary fraction of different evolutionary phases of stars, indicated by their position in the CMD, we find that the binary fractions of the sub-populations, that is MS stars, stars around the TO, BSS, as well as Be stars, are different. Given that some of these populations, in particular BSSs and Be stars, are possibly dominated by BiPs, we refrain from applying our bias correction to these groups. While the observed spectroscopic binary fraction of MS stars is about 10\%, the observed spectroscopic binary fraction around and above the cluster turnoff is significantly higher, that is around 30\%. The observed spectroscopic binary fraction of star in the Be star region is, however, only $\sim$6\%. These observational results are compatible with qualitative expectation from binary evolution and set the stage for a detailed comparison with predictions from binary population synthesis.

Comparing the overall bias-corrected close binary fraction of NGC~330 with the binary fractions determined for B stars in clusters in the LMC as well as in the Milky Way \citep{Kobulnicky2014, Dunstall2015, Banyard2021}, we find that it is generally lower. This is probably a combination of two main effects, which are difficult to disentangle. Firstly, there could be a dependence of the binary fraction on the metallicity of the host environment. Secondly, the clusters studied previously are very young (i.e., only a few Myr) and are thus more representative of the initial multiplicity conditions. Additionally, while focusing on the B-star population, previous works mostly focused on early-type B stars, while most of the stars in this study are around spectral type B5, that is they have lower masses.

In a subsequent paper, we will use the derived RVs to combine the six observational epochs for each star in order to boost the S/N ratio in the observations. These will then be used to obtain stellar parameters, such as effective temperatures, surface gravities, and particularly rotational rates. Investigating those across the CMD will allow us to further investigate the physical properties of the cluster members and to identify additional possible binary interaction products.

\begin{acknowledgements}
The authors would like to thank the referee, Max Moe, for his constructive feedback on the manuscript. The authors acknowledge support from the FWO\_Odysseus program under project G0F8H6N and from the European Space Agency (ESA) and the Belgian Federal Science Policy Office (BELSPO) through the PRODEX Programme. The research leading to these results has received funding from the European Research Council (ERC) under the European Union's Horizon 2020 research and innovation programme (grant agreement numbers 772225: MULTIPLES).
Parts of the analysis in this project are based on the python code \textsc{photutils}.
\end{acknowledgements}

\bibliographystyle{aa}
\bibliography{papers}

\begin{appendix}
\section{Two-dimensional bias correction plots}\label{App:bias_full}

\begin{figure*}[t!]\centering
    \includegraphics[width=\hsize]{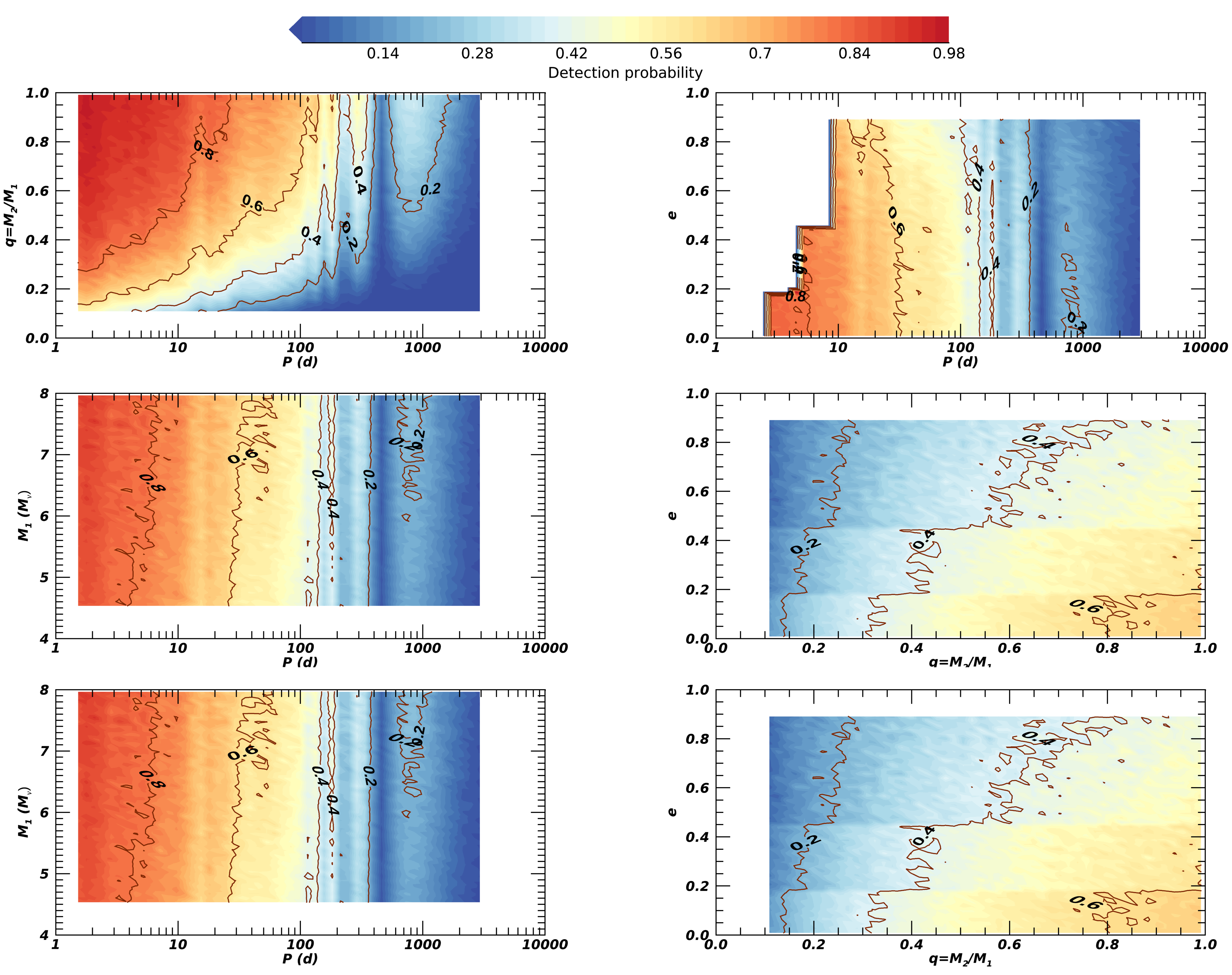}
    \caption{\label{Fig:biascorr_2D_SB1} Binary detection probability as a function of period, mass of the primary $M_1$, mass ratio $q$ and eccentricity $e$, computed for the SB1 bias. }
    \end{figure*}

\begin{figure*}[t!]\centering
    \includegraphics[width=\hsize]{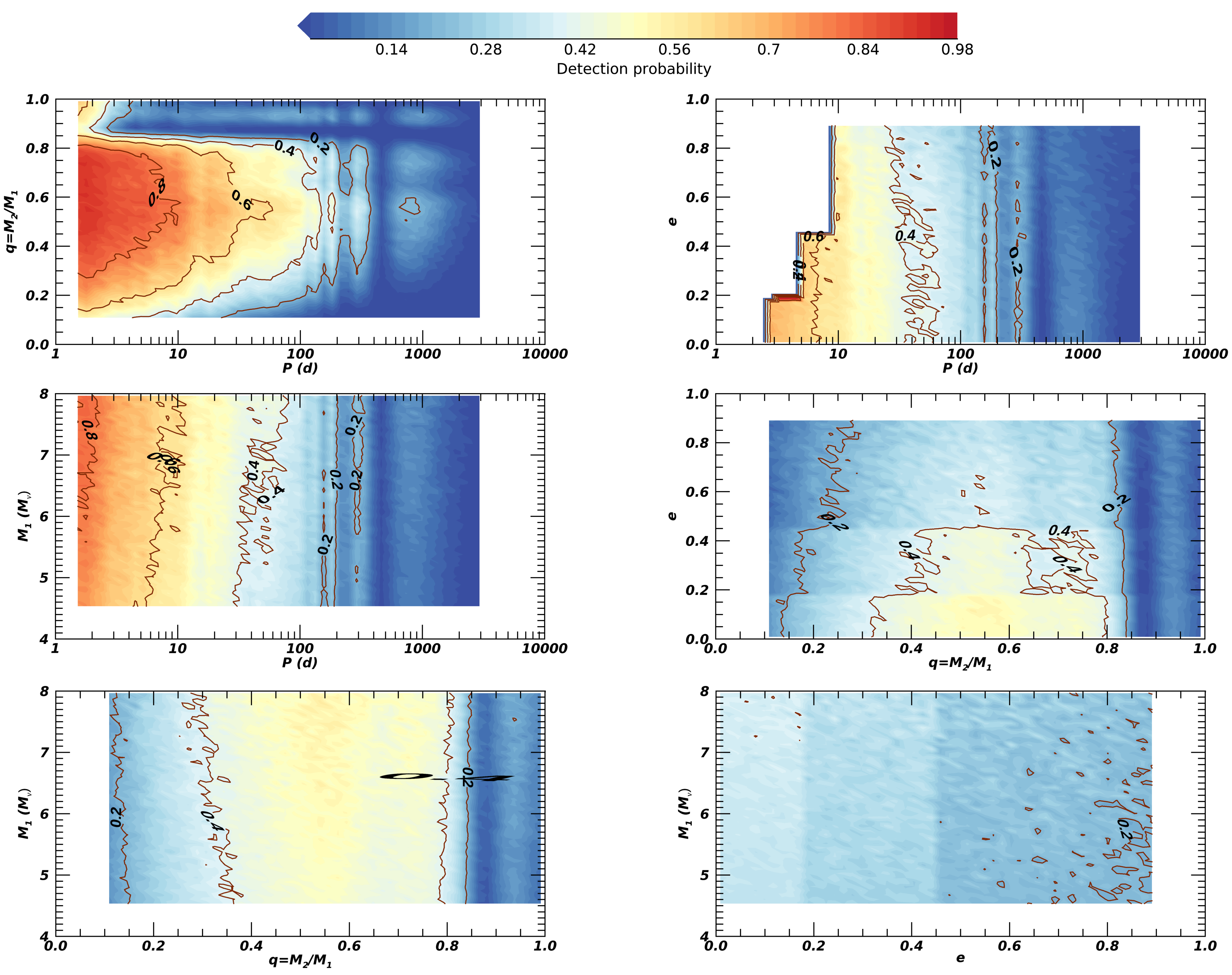}
    \caption{\label{Fig:biascorr_2D_SB2} Same as Fig.\,\ref{Fig:biascorr_2D_SB1}, but computed taking into account both the SB1 and the SB2 bias. }
    \end{figure*}

Figure \,\ref{Fig:biascorr_2D_SB1} and Fig.\,\ref{Fig:biascorr_2D_SB2} show the 2-dimensional representations of the bias correction calculated for the SB1 and as well as the SB1+SB2 bias, respectively that have been computed with flat period and mass-ratio distributions and a $\sqrt e$ eccentricity distribution. These comparison between these plots reveal the strong impact of the SB2 bias for all binaries with $q>0.8$, but for the  ones with the shortest periods. The staircase aspect of the $e-\log P$ plots results from our simplistic pseudo-circularisation prescription that aims to avoids unrealistically high eccentricity at short periods as stars in those systems that would collide at periastron.

\section{Magnitude-dependant bias correction}\label{App:bias}
\begin{figure}[t!]\centering
    \includegraphics[width=0.95\hsize]{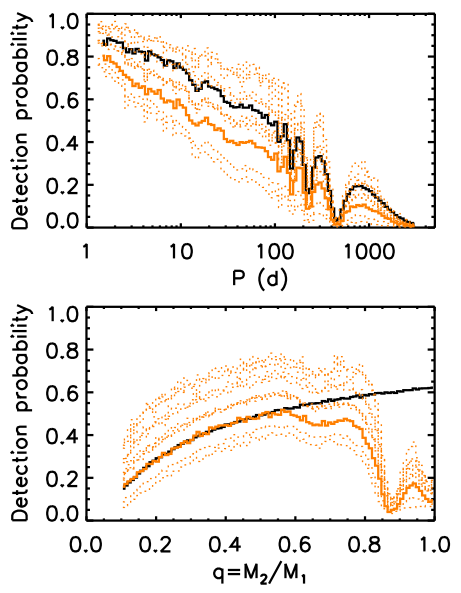}
    \caption{\label{Fig:mag_biascorr} Binary detection probability as a function of period (top panel) and mass ratio (bottom panel). The different orange dotted curves correspond to different magnitude bins, where the detection probability is highest for the brightest stars. The continuous orange line shows the overall detection probability for the full simulation while the continuous black gives the detection probability for the SB1 bias only (see Sect.\,\ref{s:sb1-bias}).}
\end{figure}

As explained in Sect.\,\ref{subsec:obsbias}, apart from the overall bias correction for the entire sample, we compute a magnitude-dependant bias correction. For this, we re-run our  simulation including the SB1 and SB2 bias for different magnitude bins between F814W = 14.5 and 19\,mag in steps of 0.5\,mag while changing the primary mass as well as typical RV errors accordingly.  The magnitude-dependant detection probability as a function of initial periods $P$ as well as mass ratios $q$ is shown in Fig.\,\ref{Fig:mag_biascorr}. The primary mass as well as the overall detection probability for the SB1+SB2 bias is given in Table\,\ref{Tab:mag_pdetect} for each magnitude bin.

\begin{table*}\centering
    \caption{\label{Tab:mag_pdetect} Magnitude-dependant bias correction. The magnitude bin is indicated in the first column with the corresponding selected initial masses in the second column. In the third and fourth column we give the number of stars with measured RVs and the observed close binary fraction in the corresponding magnitude bin, respectively. The last two columns give the binary detection probability considering the combined SB1+SB2 bias and the bias-corrected close binary fraction.}
    \begin{tabular}{lrrrlr} \hline \hline
        F814W [mag] & $M_1$ [\Msun] & N & $f_\mathrm{SB}^\mathrm{obs}$ [\%] & p$_\mathrm{detect}^\mathrm{SB1+SB2}$ & $f_\mathrm{cl}$ [\%]\\ \hline
        \hline
        14.5 - 15.0 & 8.5 - 9.0 &  2 &  0          & $0.60\pm 0.35$ & ...\\
        15.0 - 15.5 & 8.5 - 9.0 &  6 & 33 $\pm$ 19 & $0.61\pm 0.27$ & $54^{...}_{-21}$\\
        15.5 - 16.0 & 8.0 - 8.5 & 14 & 21 $\pm$ 11 & $0.53\pm 0.27$ & $40^{+44}_{-14}$\\
        16.0 - 16.5 & 7.5 - 8.0 & 26 & 23 $\pm$ 8  & $0.51\pm 0.19$ & $45^{+25}_{-13}$\\
        16.5 - 17.0 & 7.0 - 7.5 & 25 & 16 $\pm$ 7  & $0.41\pm 0.21$ & $39^{+30}_{-12}$\\
        17.0 - 17.5 & 6.5 - 7.0 & 42 & 19 $\pm$ 6  & $0.44\pm 0.15$ & $43^{24}_{-11}$\\
        17.5 - 18.0 & 6.0 - 6.5 & 60 & 10 $\pm$ 4  & $0.45\pm 0.19$ & $22^{+15}_{-6}$\\
        18.0 - 18.5 & 5.5 - 6.0 & 73 &  8 $\pm$ 3  & $0.30\pm 0.08$ & $27^{+17}_{-9}$\\
        18.5 - 19.0 & 5.0 - 5.5 & 33 &  6 $\pm$ 4  & $0.26\pm 0.18$ & $23^{+58}_{-9}$\\
        \hline
    \end{tabular}
\end{table*}

\section{Tables with radial velocity measurements}\label{app:table_rvs}
Table \ref{tab:all_params} gives an overview over the coordinates, F814W magnitudes, spectral types, the measured RVs with errors for all B and Be stars as well as BSGs, and possible comments.

\begin{sidewaystable*}
\renewcommand{\arraystretch}{1}
\caption{\label{tab:all_params} Compilation of all parameters derived for the sample stars. The first, second, and third columns give the identifier as well as coordinates. The column 'F814W' gives the F814 magnitude from \citep{Milone2018}. The column 'SpT' gives the derived spectral types. We then give an overview over the measured RVs and associated errors for each epoch. The next column indicates whether the star was flagged as binary ('-' indicates that no RV measurements were possible, 'sin' means no significant RV shifts were detected, 'SB1' means it was flagged as binary based on RV variability, 'SB2' indicates a composite SB2 spectrum). The last column gives possible comments. A full version of this table is available electronically. The first few lines are shown as an example.}
 \begin{small}
\begin{tabular}{lllllllllllll}
\\ \hline
Ident & RA & DEC & F814W & SpT & RV$_1$ & RV$_2$ & RV$_3$ & RV$_4$ & RV$_5$ & RV$_6$ & flag & c \\
BSM & [J2000]& [J2000]  & [mag] &  & [\kms] & [\kms] & [\kms] & [\kms] & [\kms] & [\kms] & &  \\ \hline
010 & 14.078913        & -72.472822 & 17.8 & B3e & 154.8$\pm$2.7 & 149.6$\pm$1.9 & 151.1$\pm$1.5 & 152.1$\pm$2.6 & 150.5$\pm$4.4 & - & sin & - \\
020 & 14.053745 & -72.472215 & 18.0 & B2  & 156.8$\pm$5.5 & 151.3$\pm$2.6 & 151.2$\pm$4.6 & 155.3$\pm$3.5 & 152.3$\pm$4.5 & 158.4$\pm$5.41 & sin & - \\
021 & 14.098794 & -72.471839 & 17.2 & B1  & 153.7$\pm$4.0 & -             & 148.8$\pm$4.9 & 158.4$\pm$3.4 & 168.1$\pm$2.9 & 143.8$\pm$3.5 & SB1 & - \\
023 & 14.071630 & -72.471714 & 17.7 & B4  & 156.5$\pm$4.8 & 156.1$\pm$4.7 & 154.7$\pm$5.7 & 161.4$\pm$4.8 & 154.6$\pm$5.6 & 156.3$\pm$4.6 & sin & - \\
025 & 14.071630 & -72.471714 & 18.4 & B6  & 178.9$\pm$21.4 & 158.5$\pm$28.0 & 133.7$\pm$25.7 & 111.0$\pm$27.4 & 198.9$\pm$33.2 & 103.5$\pm$30.6 & sin & - \\
...\\
\end{tabular}
\end{small}
\end{sidewaystable*}

\end{appendix}
\end{document}